\documentclass[twocolumn]{aastex61}

\shorttitle{DRAFT for remarks}
\shortauthors{Chabot et al.}

\begin{document}

\title{Coulomb explosion of polycyclic aromatic hydrocarbons induced by heavy cosmic rays: carbon chains production rates}

\correspondingauthor{Marin Chabot}
\email{chabot@ipno.in2p3.fr}

\author{M.Chabot}
\affil{Institut de Physique Nucl\'eaire d'Orsay (IPNO), CNRS-IN2P3, Univ. Paris Sud, Universit\'e Paris-Saclay, F-91406 Orsay, France}

\author{K.~B\'eroff}
\affiliation{Institut des Sciences Mol\'eculaires d'Orsay (ISMO), CNRS, Univ. Paris Sud, Universit\'e Paris-Saclay, F-91405 Orsay, France}

\author{E.~Dartois}
\affiliation{   Institut d'Astrophysique Spatiale (IAS), CNRS, Univ. Paris Sud, Universit\'e Paris-Saclay, F-91405 Orsay, France
}

\author{T.~Pino}
\affiliation{Institut des Sciences Mol\'eculaires d'Orsay (ISMO), CNRS, Univ. Paris Sud, Universit\'e Paris-Saclay, F-91405 Orsay, France}

\author{M.~Godard}
\affiliation{Centre de Sciences Nucl\'eaires et de Sciences de la Mati\`ere (CSNSM), CNRS-IN2P3, Univ. Paris Sud, Universit\'e Paris-Saclay, F-91405 Orsay, France}

\begin{abstract}
Cosmic Rays (CR) process the matter of the Interstellar Medium. Such energetic processing  not only modifies the interstellar matter but also injects chemical species in the gas phase. In this work, we study the effect of the CR on the astrophysical polycyclic aromatic hydrocarbons (PAH). For events in which many electrons are stripped out from the PAH by interaction with a heavy cosmic ray particle, coulomb explosion takes place and carbon chains are produced. The fragments production rates of carbon chains are of particular interest for astrophysical models. We computed PAH multi-ionization cross sections with an Independent Atom and Electron  collisional model. We introduced and used a model to predict the fragmentation pattern for the coulomb explosion. Experimental measurements on small hydrocarbons,  C$_{60}$ and PAHs were used to set confidence intervals on the calculations results. 
   The carbon chains production rates were calculated using different CR fluxes and elemental compositions, to account for the variations expected in various astrophysical environments. A range of PAH sizes and compactness were also explored. The PAH lifetime with respect to a standard interstellar CR flux (corresponding to an H$_2$ ionization rate of $\zeta \approx$ 6.10$^{-17}$s$^{-1}$) is found to be in the order of a few billion years. The production rates of interstellar carbon chains containing around 5-15 carbon atoms are in the order of few to many tens of percent of the H$_2$ ionization rate $\zeta$. The exact rate value relies on the nature of the PAH and on the CR composition.  In diffuse medium, with ten percent of the available cosmic carbon locked in PAHs, this process leads to carbon chain fractional abundances at steady state, in the range of  $10^{-15}$-$10^{-14}$, with a confidence interval of about one order of magnitude.  It reaches  $10^{-13}$ in quiescent dense clouds.  

\end{abstract}


\section{Introduction} \label{sec:intro}

In the Interstellar Medium (ISM), carbon atoms are locked  into many different structures. They form the CO molecules \citep[][]{Dame2001}, numerous highly unsaturated small carbon based molecules and long carbons chains based molecules \citep[][]{Brunken2014, Liszt2012, Gerin2011, Gupta2009, Teyssier2004, Fuente2003, Cernicharo2008,Cernicharo1996, Sakai2008, Bell1997}. Carbon atoms are also found in large molecules containing a polyaromatic structure, accordingly to the so-called PAH hypothesis \citep[][]{Allamandola1985, Leger84}, although, strictly speaking, no pure PAH has been identified yet. This class of molecules or small grains containing a few tens to a few hundreds of carbon atoms is believed to be the source of the so-called unidentified or aromatic infrared bands (respectively UIB and AIB) observed in emission \citep[][]{Leger84, Allamandola1985, van_Diedenhoven2004, 2007ApJ...657..810D, Berne2009, 2011A&A...525A.103C, 2011EAS....46.....J, Boersma2014}. To a large extent the cosmic carbon is locked into grains with a physico-chemical structure ranging from polyaromatic to hydrogenated amorphous carbon grains, a-CH, \citep[][]{Draine2015, Gadallah2013, Jones2013, Carpentier2012, Acke2010, Pino2008, Dartois2007, Sloan2007, 2004A&A...423L..33D, 2002ApJS..138...75P, 1995A&AS..113..561C}. In dense clouds, a fraction of carbon is also locked into ices \citep[][]{Boogert2015}. The gas phase synthesis of medium size carbon based molecules (i.e. with more than 2-4 carbon atoms) by a bottom-up pure gas phase chemistry faces difficulties in matching the observed abundances \citep[][]{Cuadrado2015, Guzman2015, Pety2012, Pety2005}. These molecules may alternatively be produced by fragmenting the reservoirs of large polyaromatic molecules, and/or interstellar solids with UV photons, shocks and Cosmic Rays.

 The relevance of Cosmic Rays irradiation in the evolution of dust grains, ice mantles  has been pointed out by, e.g., \cite{Shen04, Ivlev2015, Bringa2007, Leger85,2011A&A...526A..52M}. Efforts are presently undertaken to set quantitative estimates on the desorption processes efficiencies, i.e. the release in the gas phase of molecules, under  CR impact \citep[][]{Dartois2015a, Mejia2015, Dartois2015b, Dartois2013, Seperuelo2010, Godard11}.
 
  As will be seen in the article, heavy CR impact on PAH  produce highly positively charged species. The stability of these highly multicharged species has been poorly addressed yet. 
 The disruption of grains accelerated to relativistic speed has been studied \citep{Hoang15}. Collisional charging after traversing a gas column was modeled  and used to estimate their mean free path in ISM. 
 Destruction rates of PAH by CR has been estimated  by \cite{2011A&A...526A..52M}. In none of these works, the result of the coulomb explosion i.e. the fragment production rate, was studied. This is the main purposes of the present work.  
 
In this article we report on the coulomb explosion of highly positively charged PAHs induced by the interstellar CR leading to the formation of long carbon chains. In a first section dedicated to model issues, we recall basic concepts in collision physics. We then present a model allowing to calculate the PAHs multiple ionization cross sections.  In the same section, we introduce some basic concepts about the relaxation of finite systems. We then present a model to derive fragmentation patterns resulting from the coulomb explosion induced by CR. Experiments are used to give confidence interval of both models.  The second section of the article is dedicated to the astrophysical implications. We compute the PAH fold ionization rates for different CR  fluxes distributions and different types of PAH models. We then calculate the PAH life time with respect to CR and the production rates of carbon chains. Finally, we derive first order estimates of the fractional abundances of the carbon chains induced by CR in diffuse and in dense interstellar media before to summarize and conclude.

\section{Physics of ion - grain collision}

\subsection{Collision}

\subsubsection{Basic concepts}

CR of interest are fast particles in the range (0.1 MeV/u - 10 GeV/u).  The interaction between ion and matter results from the interactions the ion with the target electrons and nuclei. For energies larger than 0.1 MeV/u, the  ion - electrons cross sections are many orders of magnitude larger than  the ion - nuclei cross sections (elastic or inelastic) \citep{2010NIMPB.268.1818Z}. The interaction of CR with matter  is then dominated by the ion-electron interactions. For energy larger than 10 GeV/u both ion - electrons cross sections and CR fluxes are very low.       

In an ion-atom collision,  the target electrons feeling the coulomb perturbation of the ion passing through can be emitted to the continuum (ionization process), promoted to an atomic electronic excited state (excitation process) or  captured to a bound state of the fast ion (capture process). At high velocity, the ionization cross section is twice the excitation cross section, and the capture process is many orders of magnitude smaller \citep{1996NIMPB.107...71V} . For each of these processes, the fast ion transfers energy to the electron(s).  In the ionization case, the  energy loss by the ion corresponds to the ionization potential (IP) plus the kinetic energy of the ejected electron.  For the excitation case,  the energy loss corresponds to discrete atomic lines below the IP. The mean energy loss cross section in one collision is called the stopping power (usually expressed in eV cm$^2$/atom). The relative contributions of ionization and excitation to the stopping power is  $ \approx 80/20$ at all velocities  \citep{2010NIMPB.268.1818Z}. It is weakly depending on the type of atom. 
From the target atom point of view,  the result of the collision  is the sudden production of an excited or/and ionized species (the collision duration is in the order of $\approx10^{-17}$s, that is much shorter than any electronic motion). This atom/ion will relax by emission of photon (with timescale in the order of $\approx10^{-15}$s to $\approx10^{-10}$s depending on the transition) or by Auger electron emission in the case of hole in deep inner shells (with timescale in the order of $\approx10^{-15}$s). It is important to say that the quantity of energy relaxed by the atom/ion is not equal to the energy loss by the projectile, the difference depending on the shell of the active electron and the kinetic energy of the emitted electron. 

In ion-solid interaction, the elementary processes are the same that in ion-atom interaction. The Bragg's rule states that the stopping power of atoms and solids are identical within few tens of percent \citep{1982PhLA...88...26G}. Unlike the ion-atom situation,  the energy loss by the ion and the energy deposited in the solid are identical because the emitted electrons are stopped within the solid (with timescale in the order of $\approx10^{-12}$s). The energy deposited in the form of primary and secondary electronic excitations is rapidly ($\approx10^{-12}$s) distributed on the solid vibrations in a small volume around the track. This initial "heated volume"  can be dissipated by photon emission, thermal relaxation, but also possibly by permanent atom displacements.

In ion-grain collision, the above mentioned basic processes are conserved and the stopping power is the same that for atom or solid target within few tens of percents. The relation between energy loss and energy deposit within the grain is depending on the size of the grain. For large grains the emitted electrons cannot escape (or few of them at the ion entrance and exit of the grain) and the deposited energy is identical to the energy loss as in ion solid interaction. For small grains, electrons can escape and, as in ion-atom collision, there is no trivial relation between the deposited energy and energy loss. With a typical mean energy of the emitted electrons equal to 30 eV (this value does not strongly depends on the projectile and the target \citep{Chabot16}), the range of the electrons in solid carbon  is around 10 angstroms \citep{2006CPL...421..439O}.  The physical limit between small and large grain with respect to CR collision can be then set  to about 5 angstroms radius for a grain density of two grams per cubic centimeter(g/cc). Scaled to a PAH molecule this corresponds to a few hundreds of carbon atoms.
  
For the large grain with respect to this criterion, the electronic excitations (primary and secondary) convert rapidly ($\approx10^{-12}$ s) into vibrations. The resulting  heated grain may then evaporate atoms (or small molecules) and/or emit  photons and/or get structural and/or chemical modifications. The partition of the internal energy on these different outgoing channels (fragment formation, kinetic energy release, photon emission, structure modification,...) depends on the nature of the grain (covalent, polar, van der Waals,...) and on its size. We will focus the studies to small grains in this paper. 
   
For the small grains, the internal energy is resulting from two causes:  i) the holes induced by the ionizations in the inner (valence) shells, ii) the collisional excitation of other electrons into states below IP. For highly charged grains, the numerous coulomb repulsive electronic states  make the grain to disrupt under the so called coulomb explosion.
We will detail later the physics of this process.
   
\subsubsection{Model calculation for multi-ionization}
 The dynamics of CR - grain collisions is described in the framework of the impact parameter approximation in which the high velocity CR projectile, with an impact parameter ${\bf b}$ with respect to the grain center of mass, follows rectilinear trajectories with constant velocity ${\bf v}$. To calculate the ionization cross sections, we employ the so-called independent atom and electron (IAE) model \citep{Wohrer94} where all atoms and electrons are treated as independent. The IAE probabilities thus consist of the products of atomic probabilities. For example, with this prescription, the single ionization probability of a carbon grain made of n atoms is expressed as

       \begin{equation}
	\rm P_{{\rm ion}}^{(1)}({\bf b})= \sum_{i=1}^{n} P_{{\rm ion}}^{(1)}({{b}_{i}}) \prod_{j=1,j\neq i}^{n} (1-P_{{\rm ion}}({{b}_{j}}))
\label{singleion}
   \end{equation}

where b$\rm{_{i(j)}}$ stand for the impact parameter of the i(j)$\rm^{th}$ atom with respect to the projectile, $\rm P_{{\rm ion}}^{(1)}({{b}_{i}})$ and $\rm {{P}_{{\rm ion}}}({{b}_{j}})$ stands for the probabilities for single ionization and total ionization for atom i and j respectively. The atomic probabilities of processes  are computed assuming the independence of the electrons, i.e. in the framework of the well-known independent particle model (IPM) \citep{McGuire77}.  Valence as well as inner shell electrons are assumed to be active within the dynamics. The probabilities entering equation (\ref{singleion}), considering only carbon atoms, can thus be written as 
\small
    \begin{eqnarray}
\rm P_{{\rm ion}}^{(1)}({{b}_{i}}) = 2P_{{\rm ion}}^{(2s)}({{b}_{i}}) \times (1-P_{{\rm ion}}^{(2s)}({{b}_{i}}))\nonumber \\
\times(1-P_{{\rm ion}}^{(2p)}({{b}_{i}}))^2\times(1-P_{{\rm ion}}^{(1s)}({{b}_{i}}))^2 \nonumber \\ 
\rm +2P_{{\rm ion}}^{(2p)}({{b}_{i}})\times(1-P_{{\rm ion}}^{(2p)}({{b}_{i}})) \nonumber \\
\times(1-P_{{\rm ion}}^{(2s)}({{b}_{i}}))^2\times(1-P_{{\rm ion}}^{(1s)}({{b}_{i}}))^2 
\label{eq2}
   \end{eqnarray}
   
    \begin{eqnarray}
\rm 1-P_{{\rm ion}}({{b}_{i}}) = (1-P_{{\rm ion}}^{(2p)}({{b}_{i}}))^2\times(1-P_{{\rm ion}}^{(2s)}({{b}_{i}}))^2\nonumber \\ 
 \times(1-P_{{\rm ion}}^{(1s)}({{b}_{i}}))^2 
\label{eq3}
   \end{eqnarray}
\normalsize
   
In equations (\ref{eq2}-\ref{eq3}), $\rm P_{{\rm ion}}^{(2{\rm s},2{\rm p},1{\rm s})}$ refers to the probability of ionizing one electron from the $2{\rm s}$,  $2{\rm p}$ and $1{\rm s}$ sub-shells of the C atoms. In equation (\ref{eq2}), we do not take into account single ionization of the 1s electrons, since in this case, a rapid Auger deexcitation takes place and contributes to the occurrence of a double ionization \citep{Beroff12}.

Absolute cross sections are derived by integrating over ${\bf b}$ the IAE probabilities such as (\ref{singleion}). Since individual b$\rm _i$'s depend not only on ${\bf b}$ but also on the grain orientation  \citep{Wohrer94,Wohrer96}, we average the calculation over all equally probable grain orientations ($\theta $, $\varphi $ in spherical coordinates) following equation (\ref{eq4})

    \begin{eqnarray}
\rm \sigma_Q = \frac{1}{4\pi}\int_0^{2\pi}d\varphi \int_0^{\pi} \sin(\theta)d\theta \int_0^{\infty} P_{ion}^{(Q)}({\bf b},\theta,\varphi) d{\bf b}   
\label{eq4}
    \end{eqnarray}

The integrations were performed using a Monte Carlo method. 
   \begin{figure}
   \centering
   \includegraphics[width=8cm]{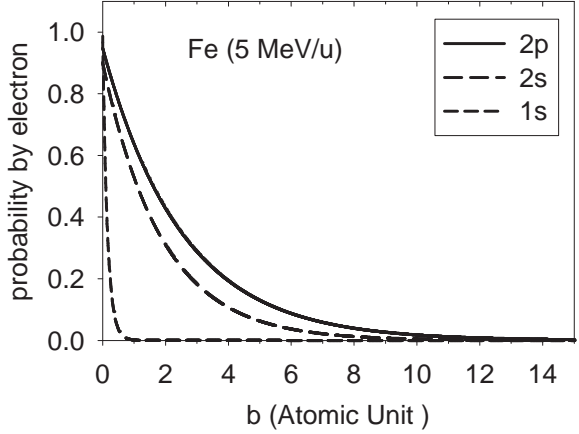}
      \caption{Carbon ionization probability $P_{{\rm ion}}^{(nl)}({{b}})$ as function of the impact parameter $b$ for the C + Fe(5 MeV/u) collision system.}     
   \label{PbFer}
   \end{figure}

The strategy used to compute the one-electron atomic probabilities, which are the basic ingredients of the IAE model, is explained below. The one-electron atomic probabilities for ionization of sub-shell nl have been obtained assuming an exponential form \citep{Kirchnner06} :

    \begin{eqnarray}
\rm  P_{{\rm ion}}^{(nl)}({{b}_{i}})= P_{{\rm ion}}^{(nl)}(0)exp(-\alpha_{nl}{{b}_{i}})  
\label{eq5}
    \end{eqnarray}
    
Taking into account the relation between the impact parameter probabilities and cross section, the factor $\rm \alpha$ is in atomic units:

    \begin{eqnarray}
\rm  \alpha_{nl} = [\frac{2P_{{\rm ion}}^{(nl)}(0)}{\sigma_{nl} / \pi a_{0}^{2}}]^{1/2}
\label{eq6} 
    \end{eqnarray}

where  $\rm \sigma_{nl}$ is the ionization cross section for an electron in the nl shell  and $\rm a_0$ is the Bohr radius. Ionization cross sections were obtained from the \cite{Kaganovich06} semi empirical fit (formula 82) using the equilibrium charge state of the projectile in H$_2$ computed with the \cite{Betz72} formula \citep{Chabot16}.  Zero impact parameter probability $P_{{\rm ion}}^{(nl)}(0)$ were computed following the so-called geometrical model \citep{Sulik87}.
The impact parameter probabilities for an iron projectile at 5MeV/u (i.e.$\approx$280 MeV) is shown  in Fig.~\ref{PbFer} as an example of such calculation.   

This model has been proved to be correct for diatomic and triatomic molecules \citep{1997PhRvL..79.1662W, 1999JPhB...32..631A, 2014PhRvA..89b2704K}, as well as for carbon and hydrocarbon molecules up to ten's of atoms \citep{2005PhRvA..72c2707M,2011PhRvA..84c2705B, Beroff12}. It has been also found  in agreement with water radiolysis  experiments \citep{2006RaPC...75..493G}.

\subsubsection{Multi-ionization cross sections for $C_{60}$}

$C_{60}$ is the typical size of PAH addressed in this paper (ie between 50 and 200 carbon atoms). It has been widely studied in collision (see for example \cite{Reinkoster01} and references therein). It will be used as benchmark for ionization but also later for fragmentation.
In addition, the typical size estimated for PAH emitters based on the PAH models \citep{2008ARA&A..46..289T} lies in this range. 

\begin{figure}
   \centering
   \includegraphics[width=8cm]{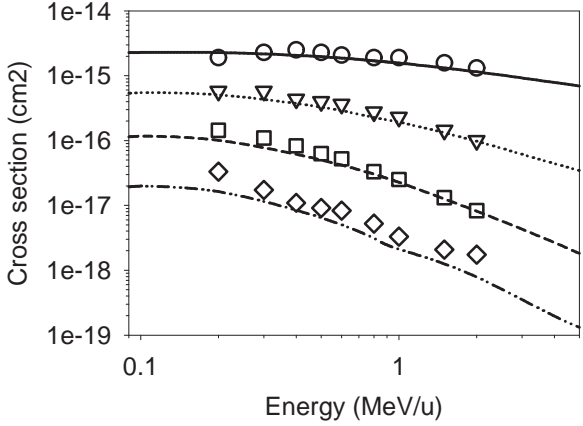}
      \caption{Single to quadruple-ionization  cross sections of $C_{60}$ by proton impact. Symbols: experimental measurements from \cite{Tsuchida98} (circle: single, triangle : double, square : triple, diamond: quadruple); lines: IAE model. }
   \label{C60pro}
   \end{figure}

\begin{figure}
   \centering
   \includegraphics[width=8cm]{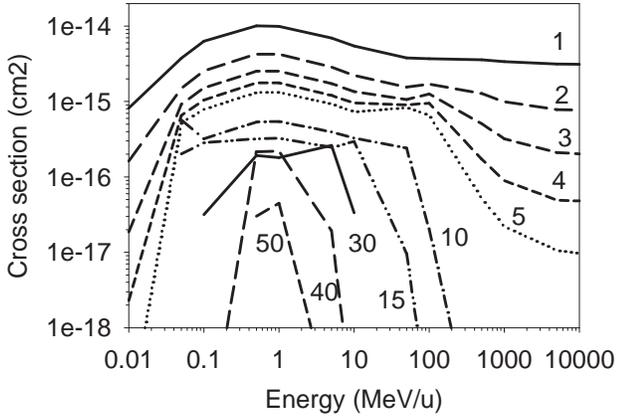}
      \caption{Single and multiple ionization cross sections of $C_{60}$ by Iron impact. The label on the curves corresponds to the ionization fold.}
   \label{C60Fe}
   \end{figure}

\begin{figure}
   \centering
   \includegraphics[width=5cm]{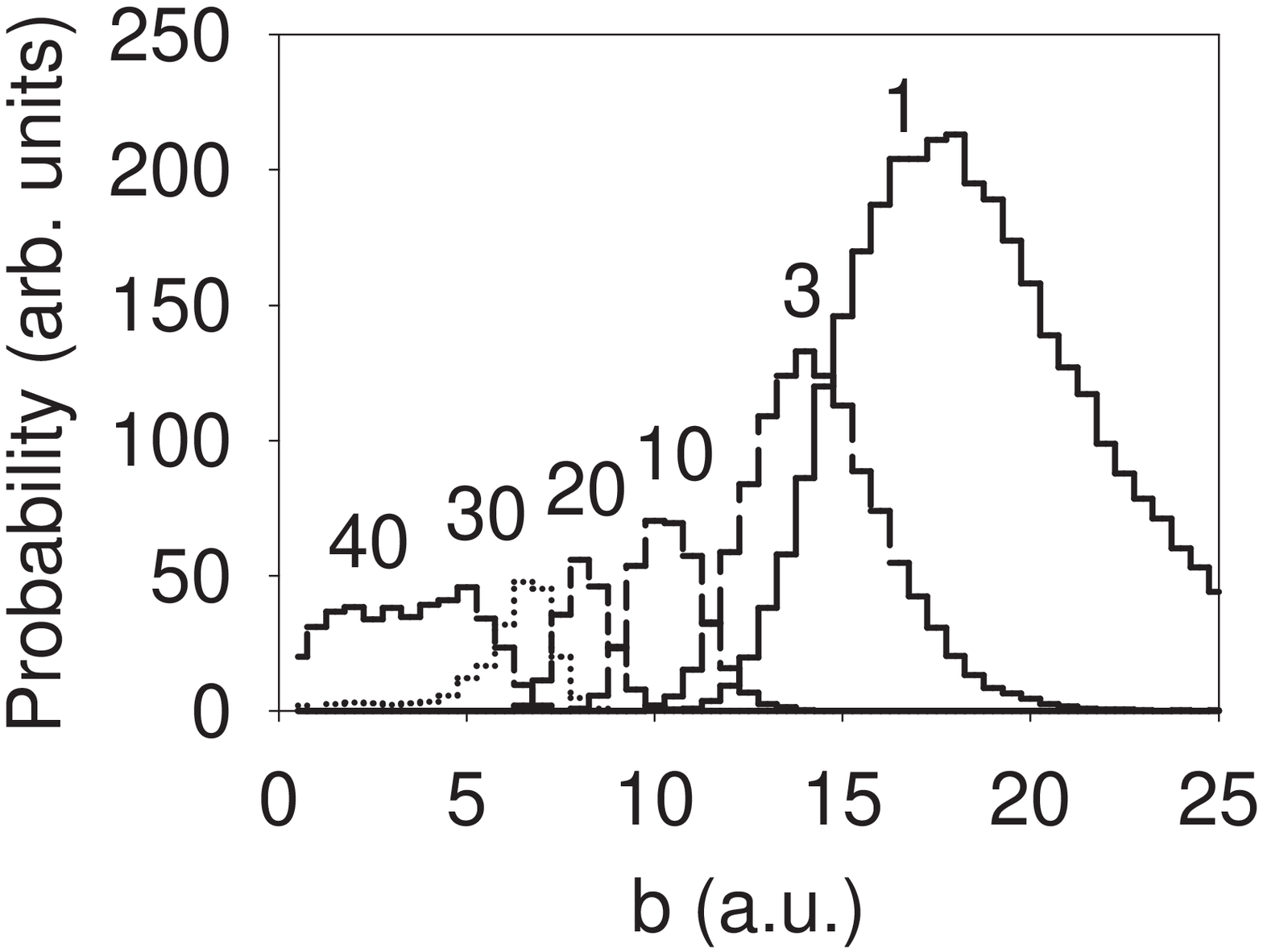}
   \includegraphics[width=5cm]{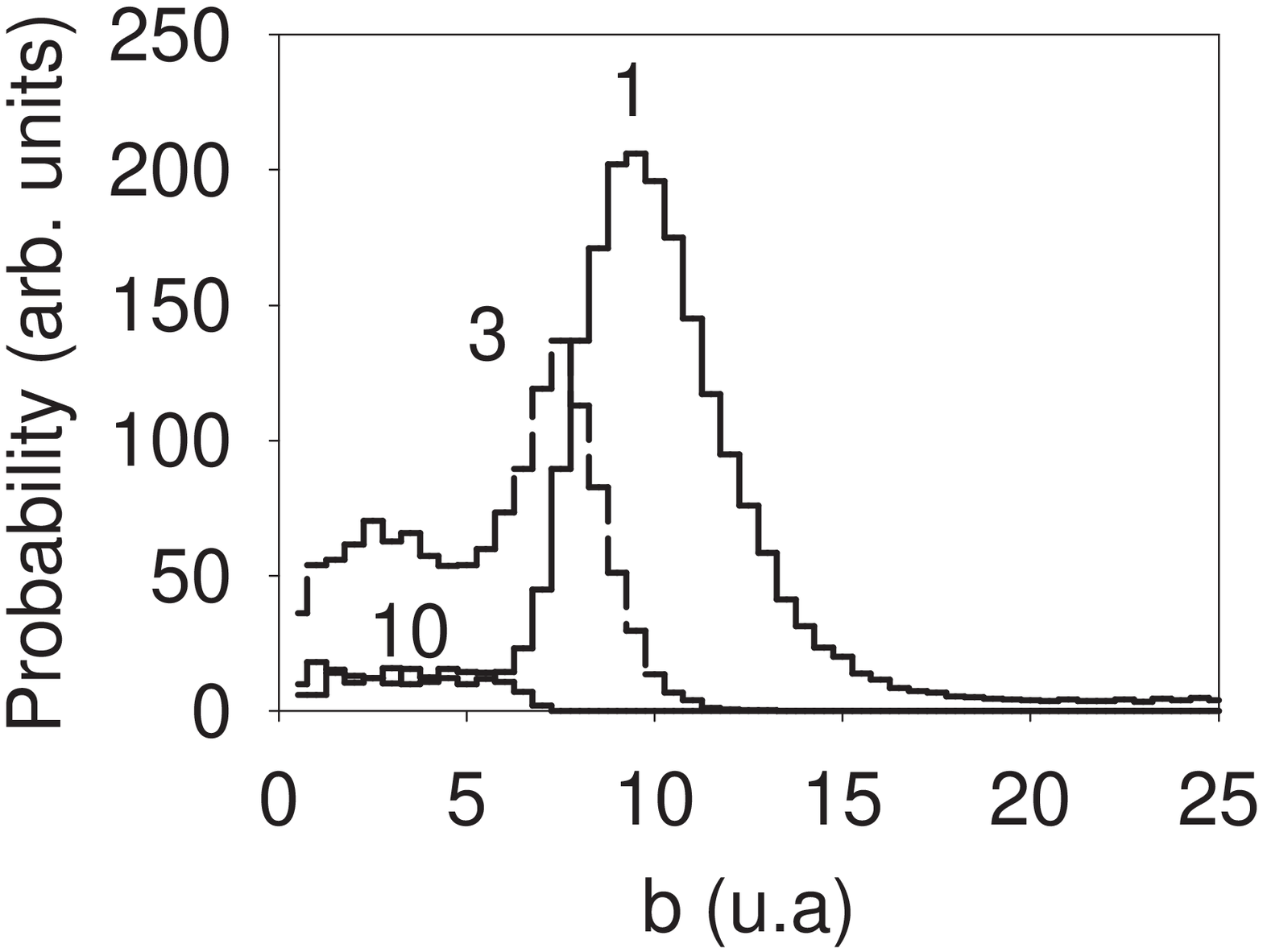}
   \includegraphics[width=5cm]{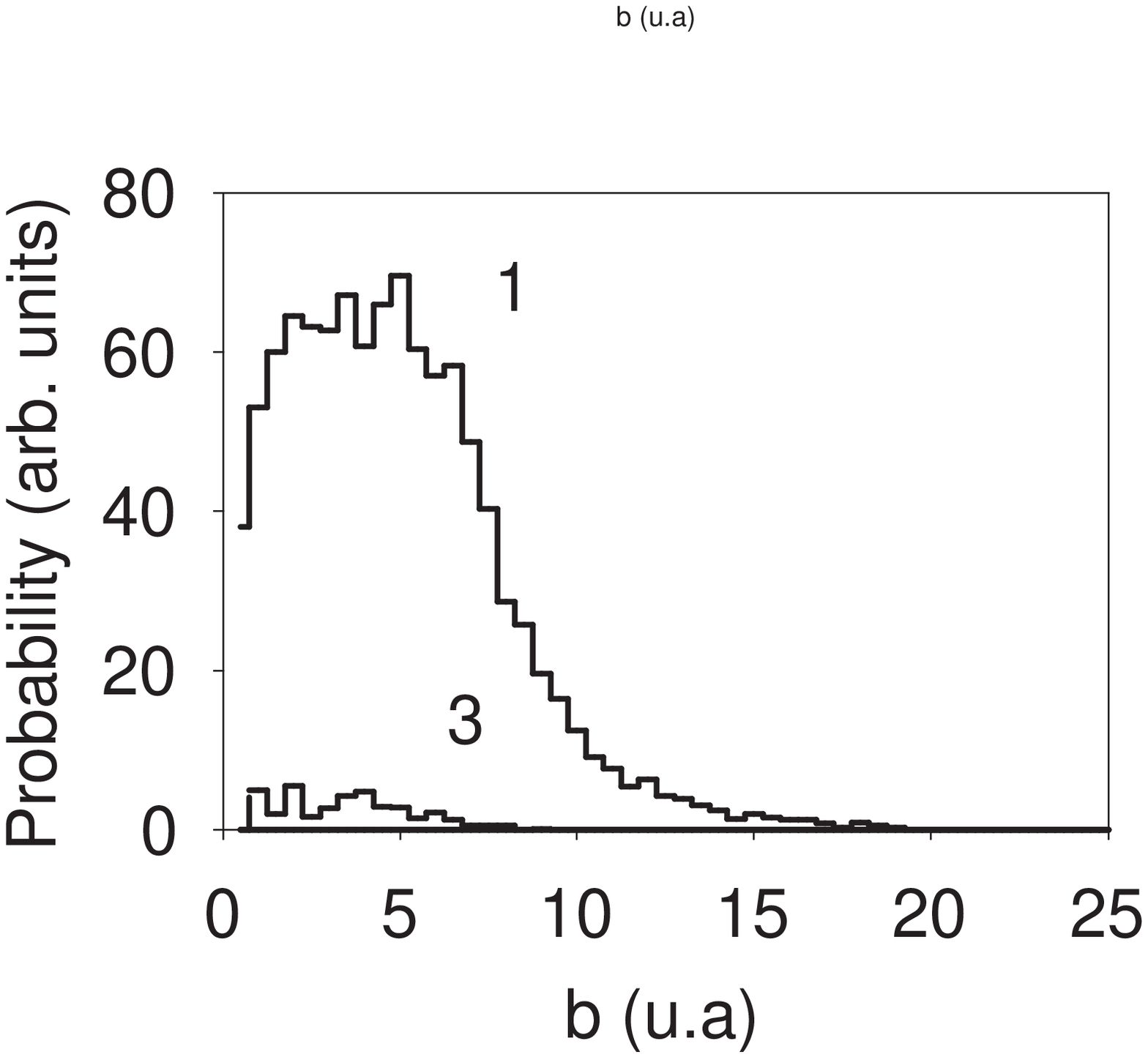}

   \caption{Relative ionization probabilities as function of the impact parameter for $C_{60}$.Upper panel: iron projectile at 1.1 MeV/u (stopping power 800 eV/$\AA$ in graphite.); middle panel: carbon projectile at 1.92 MeV/u (stopping power 100 eV/$\AA$ in graphite.); lower panel: proton projectile at 0.33 MeV/u (stopping power 10 eV/$\AA$ in graphite.).   The label on the curves corresponds to the ionization fold.}
   \label{Pb60}
   \end{figure}

The Fig.~\ref{C60pro} shows a comparison between the IAE model calculations and experimental measurements for a $C_{60}$ target ionized by protons \citep{Tsuchida98}. This measurement is the only absolute measurement that we found in the literature concerning multi ionization of large carbon structures by fast ions. The agreement between calculations and experiments is very good.  
The calculated cross sections for an iron projectile are displayed in Fig.~\ref{C60Fe}. The curves peak around 1 MeV/u. This energy corresponds to the highest stopping power for Fe projectile (800 eV/$\AA$ in graphite). It is noticeable that the low fold ionization cross sections may exceed the geometrical cross section of $C_{60}$ ($\approx 3.10^{-15} cm^2$). The probabilities of the Q fold ionization as a function of the impact parameter for $C_{60}$ are shown in Fig.~\ref{Pb60}. Three different projectiles of different stopping power (10, 100 and 800 eV/$\AA$ in graphite) have been used. There is a strong relation between the Q fold ionization  and the impact parameter extension. For the heaviest projectile, the contributing impact parameters to low Q values are much larger than the size of the $C_{60}$ molecule (6.6 atomic unit). Since cross sections are proportional to the $ b \times p(b)$ product, it explains why the cross sections exceed the geometrical cross sections for the low Q fold ionizations.

\subsubsection{Multi-ionization cross sections for PAH - role of the hydrogenation degree}
\label{1.2.4}
Interstellar PAHs  contain hydrogen.  For the ionization cross sections, this is of little importance because at first order the ionization cross sections scale as the number of valence electrons. The cross sections for a PAH  containing n carbon and m hydrogen atoms is then approximately given by:
 
    \begin{eqnarray}
\rm  \sigma({C_nH_m}) \simeq (1+\frac{m}{4 \times n}) \times  \sigma({C_n})          
\label{eq18}
    \end{eqnarray}
Even for a hydrogenated laboratory PAH such as coronene (C$_{24}$H$_{12}$), the error on the cross sections is around 10 $\%$ if  hydrogen is omitted.  We neglect hereafter hydrogen in the  collisional calculations. 

\subsubsection{Internal energy following multi-ionization}

\begin{figure}
   \centering
   \includegraphics[width=8cm]{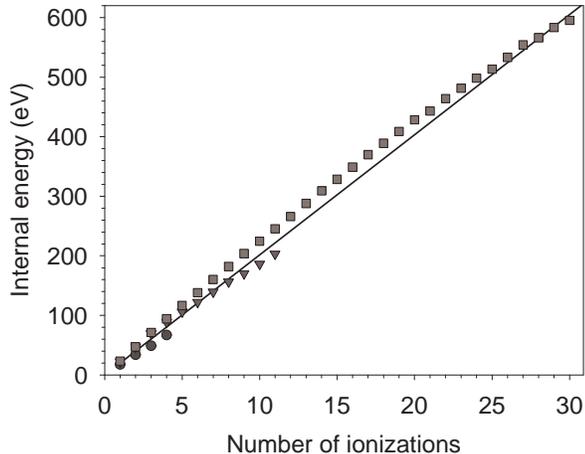}
      \caption{Internal energy within $C_{60}$ as function of the number of ionizations.  Square: Fe projectile at 5 MeV/u (500 eV/$\AA$ in graphite.); triangle: C projectile at 1.92 MeV/u (100 eV/$\AA$ in graphite.); circle: proton projectile at 0.33 MeV/u (10 eV/$\AA$ in graphite.). The line is a linear regression on all  points.}
   \label{excitC60}
   \end{figure}

To get insight into the behavior of a grain collided by a CR, not only the charge state is needed, but also its internal energy. This energy is resulting from both inner valence holes and  electronic excitations.  For small carbon clusters (n=2 to 10), the first contribution (inner valence holes)  has been calculated in \cite{Beroff09}. Dipole approximation in the collision was assumed. Then, similarly to photoionisation calculations \citep{Krummacher83},  pole strength of carbon clusters (i.e. probability of populating a level by ejecting an electron from an orbital) was used to obtain internal energy distributions. For all the sizes, the mean internal energy resulting from ionization was found to be 15 $\pm$ 2 eV. Since it is  governed by the $\sigma $ and the $\pi $ molecular orbitals of the C-C bonds, this means that the internal energy corresponding to a single ionization should apply for all sizes of carbon grains. The second contribution (internal energy corresponding to  electronic excitation) has been calculated for same carbon clusters in \cite{Chabot06}. The IAE model was used with energy differential impact parameter probabilities calculated with Classical Trajectory Monte Carlo (CTMC) theory.
The mean energy was found to be 13 $\pm$ 2 eV independently of the cluster sizes. Here again this mean energy should apply for all the carbon grain sizes.
 To derive the relative contributions of ionization and excitation to the internal energy following a Q fold ionization, we performed calculations of ionization cross sections together with the excitation. The impact parameter probability of excitation was approximated by:
  
   \begin{eqnarray}
\rm P_{{\rm ex}}({{b}_{i}}) = \frac{\sigma_{ex}}{\sigma_{ion}}P_{{\rm ion}}^{(2p)}({{b}_{i}})
\label{eq7}
   \end{eqnarray}

 with the ratio between excitation and ionization cross section $\frac{\sigma_{ex}}{\sigma_{ion}} $ taken equal to 1/2 \citep{1996NIMPB.107...71V}. The double differential cross sections with charge and number of excitations was therefore obtained. The mean internal energy for a given charge state Q was then defined by :

   \begin{eqnarray}
\rm E^*(Q)=Q\times 15 ev + N_{ex}(Q)\times 13 eV 
\label{eq8}
   \end{eqnarray}
   
where $N_{ex}(Q) $ is the mean number of excitation for a given Q fold ionization.
The Fig.~\ref{excitC60} shows this mean internal energy as function of the charge state Q for the $C_{60}$ ionized with three projectiles of different stopping powers (10, 100 and 800 eV/$\AA$ in graphite.). The slopes of the three curves are slightly different. It arises from the multiple excitation contribution that depends on the excitation probability, scaling with the stopping power. The relation between charge and mean internal energy based on these calculations, using a linear regression, writes:

   \begin{eqnarray}
\rm E^*(Q) \approx Q \times 20 ev 
\label{eq9}
   \end{eqnarray}
 
Only slight deviations from this formula are expected for different grain sizes, geometries and hydrogenation degrees.
Finally, following ionization, the coulomb potential energy has to be considered. Indeed, if many  electrons are removed only from the highest occupied molecular orbitals (HOMO), no internal energy in the form of holes will be produced but  the grain will dissipate energy by fragmenting because the produced charge species is unstable. To estimate this potential energy, the model of the point charge \citep{Sampoll92} can be used. In this model, the charges are placed at the positions of the ionized atoms and a potential coulomb repulsive  energy calculated without dynamical screening by the electrons. This model has been proved to be correct for coulomb explosion of molecules \citep{Mathur04,Siegmann00,Jana11,Beroff12}. Models of coulomb explosion with various screening functions have been also widely used in ion-solid interaction to explain sputtering  \citep{Bringa02}.   

The point charge coulomb potential energy writes:

    \begin{eqnarray}
 E_{c}(eV) = 14.4 \sum_{i=1}^{n}\sum_{j=i+1}^{n} \frac{q_i q_j}{r_{ij}(\AA)}    
\label{eq10}   
    \end{eqnarray}
with $\rm q_{i,(j)}$  the charge state of the atoms i,(j) and $\rm r_{ij}$ the relative distance between the atoms i and j.
 
For a practical point of view, we computed  this energy for each event in the Monte Carlo calculation. 

   \begin{figure}
   \centering
   \includegraphics[width=8cm]{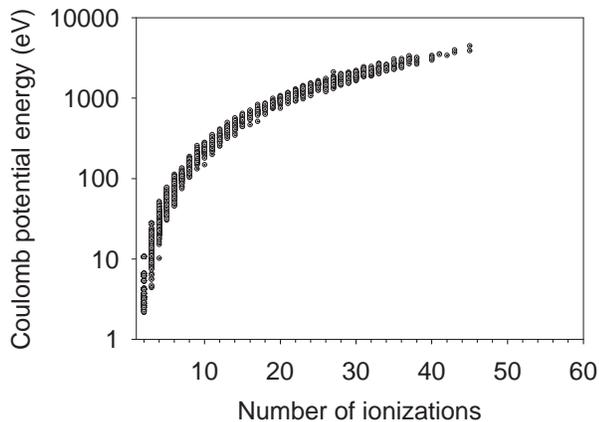}
      \caption{ Point charge coulomb potential energy, $E_c$ in $C_{60}$ as function of the number of ionizations (Fe projectile at 5 MeV/u).  Each point corresponds to one trajectory in the Monte Carlo simulation.}
   \label{Ecoul}
   \end{figure}

The Fig.~\ref{Ecoul} displays the result of this calculation as function of the charge state for $C_{60}$ ionized  by a Fe projectile at 5 Mev/u. The points dispersion, each one corresponding to an event in the calculation, reflects the varying distances between charges.\\

\subsection{Grain relaxation}

\subsubsection{Basic concepts}

The basic concepts for the fragmentation of finite systems are well understood since both nuclear and  molecular physics have documented this topic \citep[e.g.][]{ Bondorf95, Gross95, Gross97, Schapiro97, Brechignac02, Campbell96}. Unlike macroscopic systems,  microcanonical formulation (i.e. internal energy, number of constituents and charges are external quantities), is essential. The calorific curve (i.e. the microcanonical temperature as function of the internal energy)  experiences a fragmentation phase space transition (FPST). At low internal energy, the curve rises linearly, then exhibits a more or less pronounced plateau before rising again linearly. This PST corresponds phenomenologically to the opening of the particles emission channels with some enthalpy formation costs \citep{2005PhRvA..71c3202D}. Following this calorific curve, different dissipation regimes can be drawn from low to high internal energy. In the radiative regime, the internal energy is dissipated by infrared emission; in the evaporation regime, the emission of constituent cools down the system; in the fission regime, the energy is used to overcome fission(s) barrier(s); in the multi-fragmentation regime, the energy is used to break simultaneously many bonds, producing many fragments carrying kinetic energy; finally in the vaporization regime, all the internal energy goes into kinetic energy. The overlap between the various regimes is usually important. The internal energy where FPST transition occurs depends on the cohesive energy of the system, its size but also its charge. Indeed, as the charge state increases the system is less robust under coulomb forces and the FPST is occurring at lower energy. For very high charge states, since the object is highly unstable, only multi-fragmention is occurring and the FPST is disappearing. We will treat  in the following only the multi-fragmentation channels of highly charged species even if there are not the most probable species created by CR (see Fig.~\ref{C60pro} and Fig.\ref{C60Fe}). In the low charge cases, the grain behavior will be very similar to photo-ionized or photo-excited grains (with competition between IR emission and  evaporation.)

\subsubsection{Model calculation for the multi-fragmentation pattern}

\begin{figure}
   \centering
   \includegraphics[width=8cm]{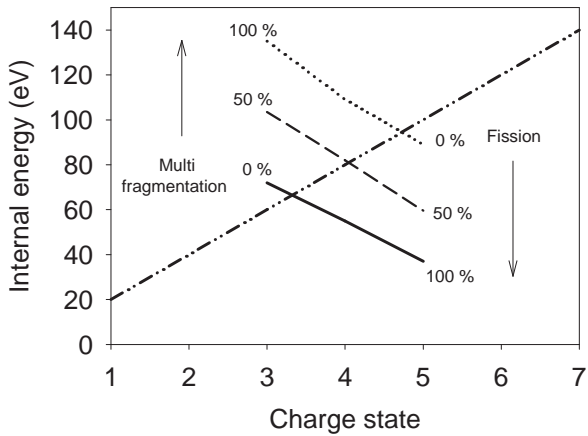}
      \caption{ Limits of the different fragmentation regimes for $C_{60}$ as function of the charge state (adapted from \cite{Qian13,Martin08}). The dashed line corresponds to equally probable fission and multi-fragmentation mechanisms. The full line delimits  the apparition of multi-fragmentation and the dotted line  the disappearance of the fission.  Percentages indicate where the mechanism appears or disappears. Arrows indicate increasing contribution  with a scale in percent. The dotted dashed line corresponds to the internal energy calculated with Eq.~\ref{eq9}}
   \label{MFC60}
   \end{figure}

The first task to model the behavior of the  ionized  and excited carbon grains under CR impact is to get the limits where the different dissipation regimes apply ( i.e. evaporation / (quasi)fission / multi-fragmentation). For that purpose, we used the fragmentation of multi-charged $C_{60}$ as prototype. In particular we used the experiments of Serge Martin team \citep{Qian13,Martin08}. In their experiments, they measured the charged fragment distributions for various charge states and, very importantly, for various internal energy bins. With mass distribution of fragments measurements, they defined limits for the different fragmentation regimes. The Fig.~\ref{MFC60}, adapted from their studies, shows the limits of multi-fragmentation for the $C_{60}^{Q+}$ ions. In the same figure, we report the internal energy resulting from the ionization by CR (Eq.~\ref{eq9}). For charge states above four,  the multi-fragmentation is the dominant relaxation mechanism. To apply these limits to any grains, we assumed that they scale as the Q/Nc ratio where Nc is the number of carbon atoms in the grain. This ratio gives somehow the number of charges (and energy) by volume unit, which is  the relevant parameter for a collision taking place in the volume. We will test later the validity of this criterion  with multi-fragmentation measurements on some laboratory PAHs.

The next task is to predict fragmentation patterns. For the low charge states, small fragments either neutral or charged may be emitted sequentially (in competition with radiative deexcitation). The probability of occurrence of a given fragment size is strongly connected to its exothermicity. For high charge states (Q/Nc $>$ 4/60), no measurement of complete fragmentation patterns for large carbon structure have been reported. Indeed measurements on large systems concern inclusive distributions (i.e. summation of a particular  fragment over many channels of fragmentation). Moreover, in all experiments, only charged fragments are observed while neutral fragments are also produced in the multi-fragmentation process. There are measurements dealing with complete patterns of fragmentation,  but only for  small multi-charged carbon clusters (C$_n^{Q+}(n=2-10;Q=0-4)$) \citep{Chabot10}. We will extend the results of these studies to small grains and use charged fragments mass distributions obtained with $C_{60}$, and PAH to check its reliability. 

 In the experiment of \cite{Chabot10},  small carbon clusters were  produced with a large  distribution of isomeric forms  and collided at  high velocity with a He atomic target at rest in the laboratory. From the complete set of fragmentation branching ratios,  a scaling law for the partitioning of internal energy   was inferred.  It stipulates that the part of the internal energy going into kinetic motion scales as the charge over mass ratio (Q/Nc). Physically this means that the higher the charges, the more the system  relax by kinetic energy release rather than by opening a new fragmentation channel.  Using the scaling law of this work  we write that the number of fragments produced by the electronic excitations and ionizations,  $N_{f_{exc}}$, is:

    \begin{eqnarray}
\rm  N_{f_{exc}} = E^*(Q)/E_{bond} \times \frac{5}{6}(1-\frac{Q}{n})
\label{eq11}
    \end{eqnarray}

where  $E^*(Q)$ is the internal excitation energy calculated with formula \ref{eq9}, $E_{bond}$ is the mean energetic cost for the production of one fragment in the multi fragmentation process, and n the number of atomic constituents.
 
For the unstable charged clusters in their fundamental states (i.e. without any internal excitation energy), additional fragments were measured to be produced in a number $N_{f_{coul}} $ equal to:

    \begin{eqnarray}
\rm  Nf_{coul} = Q-Q_H ~if ~Q > Q_H ~else ~0      
\label{eq12}
    \end{eqnarray}  
Where $Q_H$ is the highest charge that the grain can support.  
Then the total number of fragments is expressed as: 
    \begin{eqnarray}
\rm  N_{f} = N_{f_{exc}} + N_{f_{coul}}       
\label{eq13}
    \end{eqnarray}

The last task of the model is to get the mass and charge distribution of the fragments. From a dynamical point of view, small fragments (1,2,or 3 atoms) are likely to be produced either at the first stage of the multi-fragmentation  or/and by the hot fragments evaporation. Based on the presence of both small and large charged fragments in experiments \citep{Qian13,Martin00, 2013PhRvA..88e2707M, 2011PhRvA..83b2704L, 2010ApJ...708..435P}, we stated that half of the fragments will be of the type of small size fragments and the other half made of large fragments. The mean size of these large fragments, $S_{Lf}$, is  obtained by:

    \begin{eqnarray}
\rm  S_{Lf} = \frac{n-\frac{N_f}{2} \times S_{Sf}}{\frac{N_f}{2}}        
\label{eq14}
    \end{eqnarray}

     with $S_{Sf}$ the mean size of the small fragments in number of constituents equal to two, and n the total number of constituents.    

\subsubsection{Model prediction for $C_{60}$}

\begin{figure}
   \centering
   \includegraphics[width=8cm]{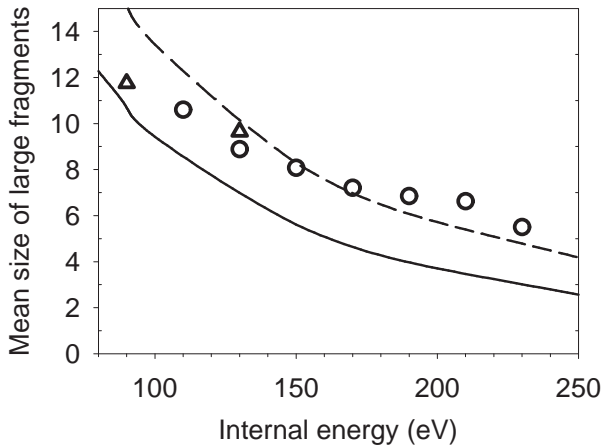}
      \caption{ Mean size in number of carbon atoms of large fragments  for  $C_{60}^{4+}$ as a function of the internal energy. Full line: result of the model with  $E_{bond}$ =7.4 eV, dashed line : $E_{bond}$ =10 eV (see text). Symbols: experimental mean size of large charged fragments. Triangles: adapted from \cite{Qian13}, circles : adapted from \cite{Martin08}}.
   \label{frag60}
   \end{figure}

To test the model reliability, we applied it first to $C_{60}$. $C_{60}$ is a very stable object because all carbons are covalently bounded to three neighbors. To reflect this stability,  we adopt for the mean energy needed to produce a fragment, $E_{bond}$ entering in formula~ \ref{eq11}, the vaporization heat of the bulk graphite (7.4 eV  www.chemicol.com/elements/carbon.html). This value is larger than typical values for carbon based molecules ($\approx$ 6 eV for the C-C bonds). The highest  charge state  that a $C_{60}$ can support, $Q_H$, entering in equation \ref{eq12}, is 14 \citep{Diaz05}. For the small fragments size, $S_{Sf}$, entering in equation \ref{eq14}, we take it equal to two. Indeed, in the experiments, $C_3$ and $C$ are the main observed  small species when multi-fragmention occurs. The Fig.~\ref{frag60} displays the mean size of the large fragments $S_{Lf}$ for the $C_{60}^{4+}$ as a function of the internal energy. In the same figure, we report experimental measurements of the mean size for charged fragments. Inclusive spectra of \cite{Qian13} (Fig.2), and \cite{Martin08} (Fig.11) were used to get these values.
For the low internal energy (i.e. less than 100 eV), the agreement of the model with the experiment is satisfactory, the mean size being underestimated by about 10 / 20 $\%$. For large internal energy, the model fails by underestimating the mean fragment size by a factor of two at 250 eV. If the mean energy needed to produce a fragment, $E_{bond}$  is increased by 25 $\%$ (10 eV), the agreement is then satisfactory. The uncertainties on this quantity will be used in the astrophysical section to set the confidence interval of the model.

\subsubsection{Model prediction for PAH}
\label{1.2.5}

Many works have been recently devoted to ion-PAH collision in the keV energy range. We test our Q/Nc multi-fragmentation criterion in light of these experimental results. In the pioneering work of \cite{2010ApJ...708..435P},  multi-fragmentation of anthracene molecules (C$_{14}$H$_{10}$) induced by 10-15 keV protons and 8-30 keV He$^{++}$ has been demonstrated through the recording of abundant low mass (2 $<$ n $<$ 15) C$_{n}$H$_{m}^+$ hydrocarbon fragments. A similar result was obtained in the experiments of \cite{2011PhRvA..83b2704L}  dealing with collisions of 30 keV He$^{++}$ with coronene (C$_{24}$H$_{12}$) and pyrene (C$_{16}$H$_{10}$) molecules. For these systems, involving low stopping power projectiles ($\approx$ 10 eV/$\AA$ in graphite), single ionization of the molecule dominates and we calculate a mean molecule charge Q$_{eff} \approx$ ~1.2. The Q$_{eff}$/Nc ratios are therefore equal to 0.085, 0.075 and 0.05 respectively for the anthracene, pyrene and coronene molecules. It is above (anthracene and pyrene) or slightly below (coronene) the  Q/Nc = 4/60 criterion. The multi-fragmentation is then expected to occur within our model for the two first cases and is close to the threshold for the third one. We note that electron transfer contributes non negligibly in the above cited experiments  whereas it is not introduced in the model, so that we can consider that overall agreement is good.
Collisions between ions and clusters of PAH have also been studied. In the first work of that kind,  clusters of anthracene were collided by 11.25 keV He$^{++}$ projectiles \citep{2010PhRvL.105u3401H}. 
The main observed deexitation channel occurs via emission of a PAH monomer rather than multi-fragmentation as in the case of the isolated PAH molecule. These results were corroborated by the work of \cite{2011PhRvA..84d3201J} in the case of 22.5 keV He$^{++}$ collisions with clusters of anthracene and clusters of coronene. Even for such large size, we can show that single ionization dominates (see bottom panel of fig. \ref{Pb60}  for the case of C$_{60}$). Then Q$_{eff}$/Nc is much lower than 4/60 and we do not expect multi-fragmentation to occur.  
We now address the question of the size and hydrogenation of the fragments. In the experiments of \cite{2011PhRvA..83b2704L},  the mean size of large fragments (i.e Nc $>$ 3) is 7.5 for C$_{24}$H$_{12}$. Applying the model with  E$_{bond}$ = 6 eV in Eq.\ref{eq12}  leads to a mean size of large fragments SLF equal to 10 while using E$_{bond}$ = 5 eV leads to  SLF=7.5. We note that we do not take into account the hydrogen content of PAH in our model. If it is perfectly justified for collisional cross sections determination, as discussed in §\ref{1.2.4}, this is not the case for fragmentation where hydrogen plays an important role. Indeed, hydrogen emission is a very probable channel because of its low binding energy as compared to C. As a consequence, it is well known that in the evaporation regime (i.e for low charge states and low internal energy) if there is hydrogen, the carbon skeleton will be preserved \citep{2003ApJ...584..316L}. Looking now at the qualitative effects that a treatment of hydrogenation would have in our model, we note that two effects play in opposite directions. First, for a given total number of fragments, hydrogen will be overrepresented in small fragments and accordingly the number of carbons to put in large fragments will be higher and so will be SLF. The second effect is related to E$_{bound}$ which will be smaller with the presence of H. It gives rise to a higher number of fragments and, as discussed before, to smaller SLF values.  A proper treatment of hydrogenation is out of the scope and the present model capability, but is of course of importance for astrophysical purposes. \cite{2010ApJ...708..435P} measured the hydrogenation degree of fragments of  C$_{14}$H$_{10}$ in their experiments. In average, 2 to 3 hydrogen were found to be linked to the carbon chains resulting from multi-fragmentation. We will consider these numbers later as "typical" in the astrophysical discussion.

\section{Astrophysical implications}

\subsection{Calculation method and inputs}

\subsubsection{Rates calculation}

The Q fold ionization rate is written:
\small
    \begin{eqnarray}
\rm  \tau_{Q} = 4\pi \sum_{Z} \int_{1 keV/u}^{10 GeV/u} j(E/A,Z)\sigma_{Q}(E/A,Z) dE/A  
\label{eq15}
    \end{eqnarray}
    \normalsize
where $\rm j(E/A,Z)$, the CR spectrum , gives the number of particles of atomic number Z and atomic mass (Z,A) of energy E per unit time, area, energy per atomic mass unit, and solid angle and  $\rm \sigma_{Q}$  is the cross section of an Q fold ionization obtained with Eq. \ref{eq4} .  We will further assume that:  
   
    \begin{eqnarray}
\rm  j(E/A,Z)= f_z ~j(E/A, Z=1)  
\label{eq16}
    \end{eqnarray}
    
where $\rm f_z$ is the fractional abundance of CR element of atomic number Z with respect to proton.

The lower limit of the integration was taken equal to 10 keV/u, and the upper limit was taken equal to 10 GeV/u. At these boundary energies both the CR spectrum and ionization cross sections are very low \citep{Chabot16}.

Following the multi-fragmentation model discussed in the previous sections of the article, the large fragment production rate is written:

    \begin{eqnarray}
\rm  \tau_{M} = \sum_Q \tau_{Q} \frac{N_f(Q)}{2} 
~~if ~\lfloor (S_{LF}(Q)) \rfloor = M    
\label{eq17}
    \end{eqnarray}

where M is the fragment mass in number of carbon atoms, $N_f$ is the total number of fragments for the charge Q  obtained with formula \ref{eq13} and $S_{LF}$ the mean mass in number of carbon atoms obtained with formula \ref{eq14}.

\subsubsection{Cosmic ray spectrum and composition}

CR are (mainly) injected into ISM by the explosion of dying stars (SNe) and confined in the Galaxy by the magnetic fields \citep[e.g.][]{2015ARA&A..53..199G, 2012APh....39...52D, 1998ApJ...509..212S}. During their motion, CR interact with the ISM hydrogen. This interaction shapes the low energy part ($<$ 100 MeV/u) of the CR spectrum  \citep{Chabot16}. 
Our local CR spectrum, measured by the voyager mission, is of this type \citep{1998ApJ...506..329W, voyager13}. This shape, called GCR spectrum in the following,  is well described by the expression of \cite{Webber83, Shen04, Godard11}. The ionization rate of H$_2$, $\zeta$ is then $\approx$ 6.3 10$^{-17}s^{-1}$ in agreement with observations in dense clouds \citep{Indriolo07}. 

In addition to GCR, local CR spectrum with a strong component of low energy particles ( i.e. $<$ 100 MeV/u) can exist  because of the proximity of a local source such as a star or a particular magneto hydrodynamical configuration. Such kind of spectrum leads to an enhanced ionization rate of H$_2$. It is proposed to be the source of the H$_3^+$ variability measured in diffuse clouds \citep{Indriolo09}. The shape of such CR spectrum can be parametrized with a E$^-1$ function below a certain energy threshold value (E$_0$=300 MeV/u) \citep{Chabot16}. $\zeta$  is then $\approx$ 1.3 10$^{-15} s^{-1}$. We will refer to  E-1 for this spectrum.

Fractional abundances of CR (entering in equation \ref{eq15}) are measured inside our solar local bubble to be very different from the solar abundances \citep{2016ApJ...831...18C, 1997ApJ...487..182M}. The heavier elements that are locked into the grains in the ISM are over represented in GCR with respect to elementary abundances. It means that at the time of acceleration both gas and grains are concerned \citep{1999JETP...89..391B,1998ApJ...500L.153L}. \cite{1997ApJ...487..197E} proposed that supernova remnant blast waves accelerate both gas and grains. The atoms sputtered from the fast moving grain by the collision with ambient gas may then be injected in the shock with suprathermal energy and further accelerated. In our local bubble, the $^{22}$Ne/$^{20}$Ne ratio in CR as compared to gas phase is increased by a factor 5  \citep{2005ApJ...634..351B}. It is interpreted as the  existence of Wolf-Rayet winds at the time of CR productions. Such kind of multi-parametric and complex mechanisms to explain the CR chemical composition makes plausible a variability in the CR composition in different particular local places of the Galaxy. To explore the influence of this parameter on our calculations, we define two elementary compositions for CR. The first one refers to a "standard" CR composition (i.e. in  our local bubble) taken from  \cite{1998ApJ...506..329W} and the second one is obtained by multiplying arbitrarily by a factor of five all species above Z=5. With this so called "heavy ions enriched composition"  and with a GCR  spectrum, the H$_2$ ionization rate $\zeta$ is 1.6 10$^{-16}s^{-1}$ . 

\subsubsection{PAH model}

Interstellar PAH  are in the range of few tens to few hundreds of carbon atoms. They are at the  origin of the unidentified aromatic infrared bands observed in emission \citep[][]{Leger84, Allamandola1985}. Nevertheless no specific species has been so far identified, at the notable exception of  $C_{60}^+$ recently \citep{2013A&A...550L...4B, 2015Natur.523..322C}. Their chemical form (aromatic versus aliphatic bonds proportion), their hydrogenation degree as well as their size and geometry are as a consequence elusive.
To cover the possible diversity of hydrocarbon carbon objects, in this size range, we used three different PAH  models.
The first one was made of the backbone (i.e. hydrogen free) of the circumcoronene ($\rm C_{54}$). 
The second one was made of a stack of 4 circumcoronene backbones with parallel orientation and inter-plane set to 3.5~$\AA$ ($\rm C_{216}$). Its radius is about 5~$\AA$ and its density is $ \approx $2 g/cc. It is more of  carbon grain type than interstellar PAH following the \cite[][]{Leger84, Allamandola1985} hypothesis (flat objects).  

The last one was made as the second one ($\rm C_{216}$), but with all  distances increased artificially by a factor two, forming an open fluffy system. Its density  is then $ \approx $ 0.25 g/cc. It is used to test the sensitivity of the results to the morphology of the objects.

In calculations, none of these PAH models possess hydrogen. For the ionization cross sections, this is of little importance as  discussed in § \ref{1.2.4}. For fragmentation it may change slightly the results as  discussed in § \ref{1.2.5}. This will be take into account in the confidence interval determination, as discussed later. 

The absolute number of interstellar PAH  experiencing coulomb explosion has to be evaluated. 
The carbon cosmic abundance is debated, but seems to be lower than previously assumed \citep[e.g.][]{Przybilla2008, Nieva2012}, with C/H of about 2.14$\times 10^{-4}$. We stated that up to 12\% of this overall carbon atoms may be locked in the interstellar PAH population \citep[e.g.][]{Draine2007, Zubko2004, Allamandola1989}. Setting all this carbon in the form of the PAH  model 2 or 3  containing 216 carbon atoms, leads to a fractional abundance of these PAHs of 1.2$\times$10$^{-7}$. For the PAH model 1, containing 54 carbon atoms,  the fractional abundance is  4.8$\times$10$^{-7}$.   

\subsection{Results and discussion}

\subsubsection{Ionization rates}

\begin{figure}
   \centering
   \includegraphics[width=8cm]{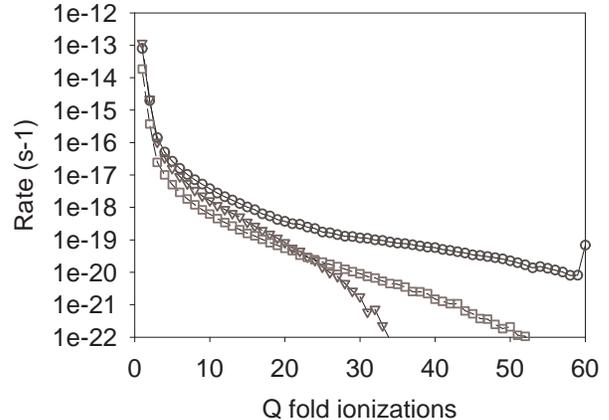}
      \caption{Rates of Q fold ionizations in a GCR spectrum with standard composition ($ \zeta$=6.3 ~10$^{-17}$ s$^{-1}$) for the different PAH  models. Squares: model 1 (Nc=54), circles: model 2 (Nc=216; d=2 g/cc), triangles:  model 3 (Nc=216; d=0.25 g/cc). At charge state 60, the sum of the rates for Q above 60 is reported.} 
   \label{tauxQ}
   \end{figure}

The  Fig.~\ref{tauxQ}  shows the rates of the Q fold ionization $\tau_{Q}$ calculated with Eq. \ref{eq15}  for the three PAH  models and for GCR spectrum with standard composition. The Q fold ionization rates for the small PAH model 1 (Nc=54) are always lower that the rates for the large PAH model 2 (Nc=216, 2 g/cc). At low charge states, the ratio is close to the ratio between the number of carbon atoms (1/4), while for high charge states it is much lower. Making the comparison between the PAH  model 2 (Nc=216, 2 g/cc) and 3 (Nc=216, 0.25 g/cc), it appears that dense PAH are more multi-ionized that fluffy PAH.

\begin{figure}
   \centering
   \includegraphics[width=8cm]{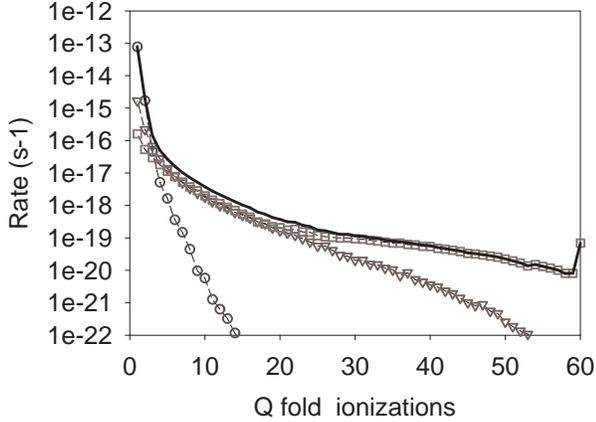}
      \caption{Contribution of the different CR species to the rates of Q fold ionization of PAH  model 2 (Nc=216) in a GCR spectrum with standard composition ($ \zeta =6.3 10^{-17} s^{-1}$). Circles : contribution of light CR ( Z=1 to 4), triangles: contribution of medium CR (Z=5 to 15), squares: contribution of heavy CR ( Z=16 to 26). The full line is the sum of all contributions.} 
   \label{tauxQZ}
   \end{figure}

For the PAH  model 2 (Nc=216), the  fig.~ \ref{tauxQZ} shows the contributions to the Q fold ionization rates of the different chemical species for an incident GCR spectrum with a standard composition. Light species (Z = 1 to 4)  dominate the low fold ionization.  Intermediate species (Z = 5 to 15) and heavy species (Z=16 to 26) have identical contributions for medium fold ionization while only heavy species contribute to high ionization folds.    

\begin{figure}
   \centering
   \includegraphics[width=8cm]{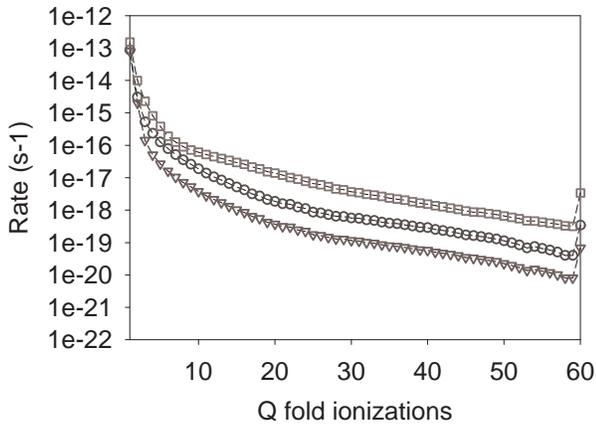}
      \caption{Rates of Q fold ionization of PAH  model 2 (Nc=216) in different CR types. Triangles : GCR spectrum with standard composition; circles : GCR spectrum with heavy enriched composition; squares: E-1 spectrum with standard composition.} 
   \label{tauxQall}
   \end{figure}

For the same PAH  model 2, the Fig~\ref{tauxQall} shows the rates of the Q fold ionization for the three assumed different types of CR. Apart from very low Q fold, the rates in a GCR spectrum are 5 times higher with heavy enriched composition than with standard composition. This is twice higher than the ratio of H$_2$ ionization rates $ \zeta$ (1.6 10$^{-16}$/6.3 10$^{-17}$). With a standard composition the E-1 spectrum is producing between 25 and 30 times more ionization than the GCR spectrum. This is a bit more than  the ratio of H$_2$ ionization rates $ \zeta$ (1.3 10$^{-15}$/6.3 10$^{-17}$). For the other PAH  models the effects of the energy spectrum and composition are the same.

\subsubsection{Life time}

The life time of PAH  in ISM is governed by the UV fields, the bath of (hot) thermal electrons, atoms and ions, the shocks and the CR \citep{2013RvMP...85.1021T, 2008ARA&A..46..289T}. The life time of PAH  under UV photons is size dependent. Indeed, above a given size, the number of modes to distribute the internal energy over  can be large enough to make thermal evaporation unprobable as compared to IR emission \citep{2003ApJ...584..316L, 2016A&A...595A..23A, 1996A&A...305..602A, 2013A&A...552A..15M}.  
For bare carbon grains, this UV limit of existence seems to lie around 30/40 atoms in the diffuse medium \citep{1995IJMSI.149..321B}. For PAH, containing  hydrogen, this limit is lower because equilibrium between hydrogen loss channels and  refill may occur without (with few) carbon  skeleton modification \citep{2004A&A...427..179H}.
The typical lower size of stable interstellar PAH under UV is admitted to be around 20/30 carbon atoms \citep{2013A&A...552A..15M, 2006A&A...460..519R}. The life time induced by electron recombination on cationic species has been found always small as compared to photodissociation and this is also the case for reactions with atoms and ions. Regarding shocks, the life time  has been studied  and found to be very long (i.e larger that 100 Myr) \citep{2010A&A...510A..36M}. For regions with very high electron temperature ($ > 10^6$ K), the PAH  lifetime may be very short (i.e. less that 1000 years) \citep{2010A&A...510A..37M}. The lifetime under CR has been estimated by \cite{2011A&A...526A..52M} to be in the order of few 100 Myrs.  In this calculation, the internal energy of PAH was set equal to an arbitrary part (0.7) of the  CR projectile energy loss. Relaxation by an evaporation model was then considered to get the destruction rates.

As we have seen in section 2, the stability of the PAH after CR impact is defined by the ratio Q/Nc . Above threshold (Q/Nc)$_{MF}$ value (4/60), the grain is destroyed by multi-fragmentation. Below this limit the overall grain stays quasi intact since the evaporation concerns mainly hydrogen atoms and with a lower probability small fragments emission.  The life time of the PAH then reads :

     \begin{eqnarray}
\rm  T_{life} ~(Myr) = ~3.16~10^{+13} /\sum_{Q=(Q/Nc)_{MF} \times Nc}^{\infty}  \tau_{Q}    
\label{eq19}
    \end{eqnarray}

Where (Q/Nc)$_{MF}$ is the ratio above which the PAH disrupts (take equal to 4/60),  $\tau_{Q}$ are the Q fold ionization rates of EQ. \ref{eq15}, and the pre-factor is to convert second in Myr.

\begin{figure}
   \centering
   \includegraphics[width=8cm]{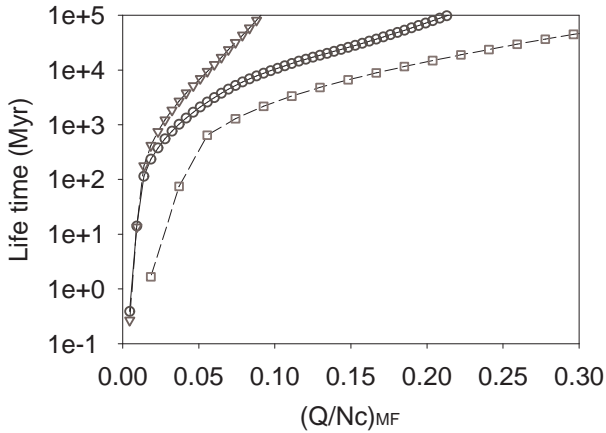}
      \caption{PAH  life time in GCR spectrum with standard composition ($ \zeta=6.3 10^{-17} s^{-1}$) as function of the ratio (Q/Nc)$_{MF}$ where multifragmentation occurs (see text). Squares: model 1 (Nc=54), circles: ~model 2 (Nc=216; d=2 g/cc), triangles: model 3 (Nc=216; d=0.25 g/cc)} 
   \label{tauxQ_lifetime}
   \end{figure}

The figure \ref{tauxQ_lifetime} shows the life times of the three PAH  models as function of the $(Q/Nc)_{MF}$ ratio under an irradiation by a GCR spectrum with a standard composition. The life time of small  PAHs is shorter than the life times of large  PAHS, and dense  PAHs are more efficiently destroyed by CR than fluffy PAHs.  At $(Q/Nc)_{MF}$ equal to 4/60 the life time is of billions of years for PAH  model 1, few billions years for PAH  model 2 and few ten's of billions year for the fluffy PAH  model 3. Note that, in fluffy  PAHs, weak bonds are likely to exist.  In this case,  (Q/Nc)$_{MF}$ corresponding to fission (found to be 2-3/60 in C$_{60}$) may be more adequate. The life time would then be reduced by an order of magnitude.

\begin{figure}
   \centering
   \includegraphics[width=8cm]{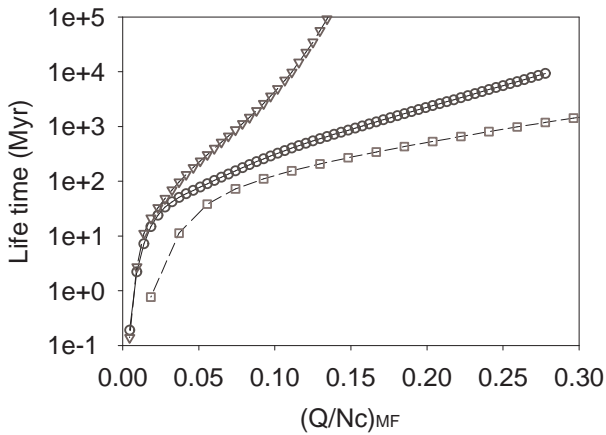}
      \caption{Same legend as Fig.~\ref{tauxQ_lifetime} for E-1 CR flux ($ \zeta=1.3 10^{-15} s^{-1}$).}
   \label{tauxQ_lifetime2}
   \end{figure}

The figure \ref{tauxQ_lifetime2} shows the life times for the three PAH  models under irradiation with an E-1 CR spectrum and a standard composition. The life time is shorter than with a GCR spectrum by about a factor of 20.  At $(Q/Nc)_{MF}$ equal to 4/60, the life time of the small PAH model 1 is around 100 million years. Even such  strong fluxes of ionizing CR  (which are moreover certainly associated with transient phenomena) are inefficient in destroying interstellar PAH within clouds life time. 

In all cases, in accordance with \cite{2011A&A...526A..52M} conclusions,  CR have little consequences on the interstellar PAH  at the time scale of astrophysical clouds in Galaxy (millions to tens of millions years). The same conclusion can be drawn for the Galactic halo where CR flux is weaker.
In the galactic center recent observations indicate that $\zeta$ could be  as high as $10^{-13}$ s$^{-1}$ \citep{2016A&A...585A.105L}. In such a place, for GCR with standard composition, the interstellar PAH life time, with respect to CR,  would be  in the order of few millions years.

\subsubsection{Fragment production rates}

\begin{figure}
   \centering
   \includegraphics[width=8cm]{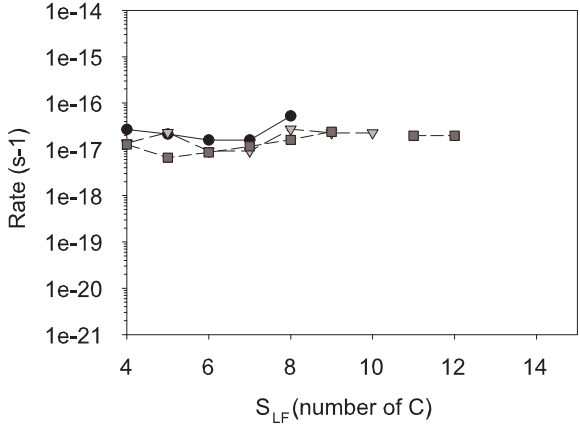}
      \caption{Fragment rates produced by GCR spectrum with standard composition ($ \zeta=6.3 10^{-17} s^{-1}$) as a function of the fragments size for the PAH  model 1 (Nc=54). Circles: $E_{bond}$=6 eV, triangles: $E_{bond}$=7 eV, squares: $E_{bond}$= 8 eV (see text). } 
   \label{fragCC_GCR}
   \end{figure}

\begin{figure}
   \centering
   \includegraphics[width=8cm]{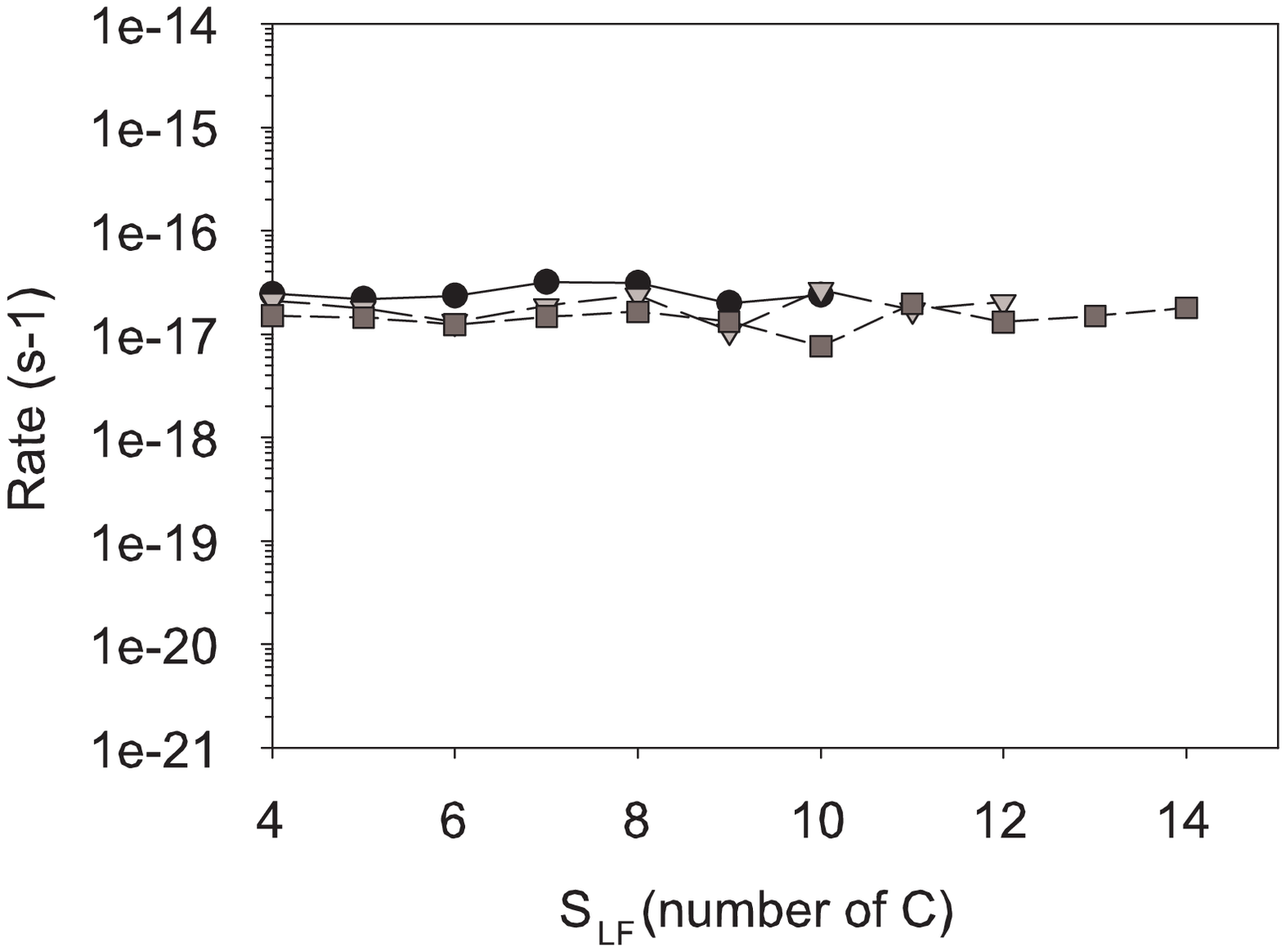}
      \caption{Same legend as Fig.~\ref{fragCC_GCR} for PAH  model 2 (Nc=216; d=2 g/cc). } 
   \label{frag4CC_GCR}
   \end{figure}

\begin{figure}
   \centering
   \includegraphics[width=8cm]{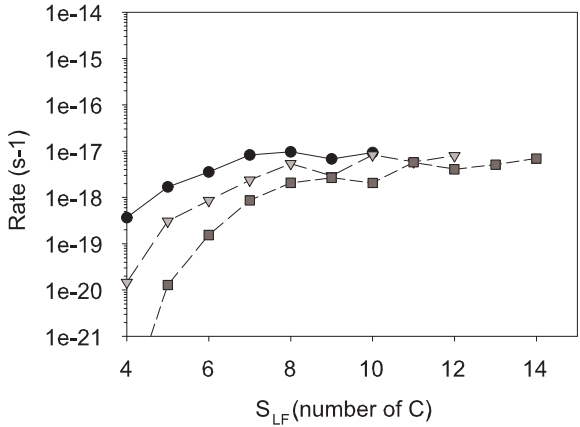}
      \caption{Same legend as Fig.~\ref{fragCC_GCR} for PAH  model 3 (Nc=216; d=0.25 g/cc). } 
   \label{frag4CCx2_GCR}
   \end{figure}

As a result of PAH destruction by CR, fragments are produced. The figures \ref{fragCC_GCR} to \ref{frag4CCx2_GCR} display the fragment production rates using Eq. \ref{eq17} for the three PAH  models irradiated by a GCR spectrum with a standard composition. The charge to mass ratio above which the PAH disrupt $(Q/Nc)_{MF}$ has been taken equal to 4/60. 

In the fragmentation calculation (see Eq.\ref{eq11}), an important parameter is the mean energy to produce one fragment, $E_{bond}$. To include the error made by a wrong determination of this number, which may depend on the PAH nature (aromaticity, hydrogenation degree, ..), we performed the calculations for $E_{bond}$ = 6, 7 and 8 eV. In all the PAH  models, a small $E_{bond}$ leads to large fragments production rates but on a smaller size interval. The fragment rates changes due to this parameter  are within a factor of 2 or 3,  except for the case of the PAH  model 3, at low fragment sizes (see Fig. \ref{frag4CCx2_GCR}). 
Excluding the fluffy PAH  model 3, it is remarkable that all fragment sizes are produced with almost identical rates. Moreover, the rates are not sensitive to the PAH size. But increasing the PAH size increases the maximum fragment size (compare Fig. \ref{fragCC_GCR} and \ref{frag4CC_GCR}).  For the fluffy PAH  model 3, the rates are always lower than the rates for the dense PAH  model 2.  There is a factor 3 between masses 15 and 10, an order of magnitude for mass 6, and more than two orders of magnitude for masses below mass 5. As we already saw for the life time, fluffy PAHs are more resilient to CR and produce less fragments. Nevertheless, as mentioned before, such fluffy PAHs are certainly subject to fission that we do not take into account in our model.  Calculations made with $(Q/Nc)_{MF}$ = 3/60 instead of 4/60 show an increase of the maximum mass by about four carbons, with equal rates. 

\begin{figure}
   \centering
   \includegraphics[width=8cm]{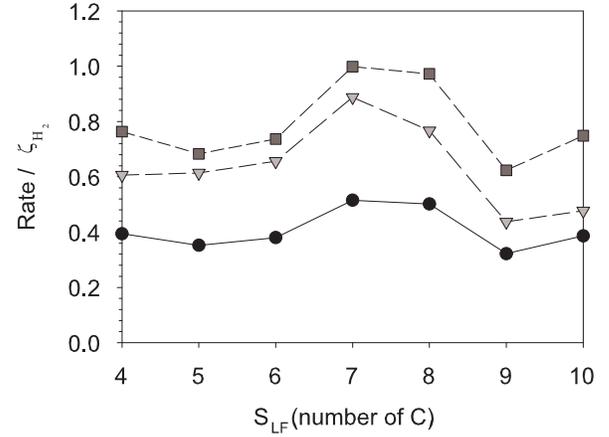}
      \caption{Production rates of fragments, in unit of H$_2$ ionization rate $\zeta$, as a function of large fragment size $S_{LF}$ for the PAH  model 2 ( Nc=216, d= 2g/cc) in various CR fluxes. Circles: GCR spectrum with standard composition ($\zeta$= 6.3 10$^{-17}$ s$^{-1}$); triangles : E-1 spectrum with standard composition ($\zeta$= 1.3 10$^{-15}$ s$^{-1}$); squares: GCR spectrum with heavy enriched composition ($\zeta$= 1.6 10$^{-16}$ s$^{-1}$).}  
   \label{frag_rel}
   \end{figure}

To better appreciate the effects of the different CR types on fragment production, we report on Fig.~\ref{frag_rel} the fragment production rates normalized to the H$_2$ ionization rates $\zeta$. The calculations were performed  for the PAH  model 2  with $E_{bond}$ = 6 eV and $(Q/Nc)_{MF}$ = 4/60. For a GCR spectrum with standard composition, the production rates are  $\approx$ 40 $\%$ of $\zeta$, whereas it is $\approx$ 60 $\%$ of $\zeta$ for a E-1 spectrum  and $\approx$ 80 $\%$ of $\zeta$ for a GCR spectrum with heavy enriched composition.

	The fragmentation process produces large fragments but also small ones. In the model, they are assumed to be made of one, two or three carbon atoms. The production rates of  these small fragments are given  in Table~\ref{smallfrag} for the different PAH  models and CR types. As for the large fragments, large and dense PAHs produces more small fragments than small or/and fluffy PAHs. GCR spectrum  with heavy enriched composition produces always more small fragments than the other CR types. Because we assumed in the model an equal population of large and small fragments, those rates are equal to the sum over the size of the large fragments production rates.   
	
\begin{table}
\begin{center}
\small
\begin{tabular}{c c c c}
\hline
 & & & \\
\textbf{PAH  model} & \textbf{Type of CR}  &  \textbf{Rate} (s$^{-1}$) & \textbf{Rate}/($\zeta$) \\
 & & & \\
\hline
 & & & \\

1 & GCR standard &  1.8(-16)  & 2.8 \\
 & GCR heavy &  8.8(-16)  & 5.5 \\
& E-1 &  3.5(-15)  & 2.7 \\
 & & & \\

2 & GCR standard &  2.2(-16)  & 3.6 \\
  & GCR heavy &  1.1(-15)  & 7.0 \\
  & E-1       &  7.3(-15)  & 5.6 \\
 & & & \\

3 & GCR standard &  4.0(-17)  & 0.6 \\
 & GCR heavy &  2.0(-16)  & 1.2 \\
 & E-1 &  1.3(-15)  & 1.0 \\
 & & & \\
\hline
\end{tabular}
\normalsize	    
    
\end{center}

\caption{\label{smallfrag} Total production rates of small fragments in s$^{-1}$ and in unit of H$_2$ ionization rate $\zeta$ (two last columns), for the three PAH  model (first column)  and different CR type (second column). A(B) = A$\times$10$^{B}$.}
\end{table}

\begin{scriptsize}

\end{scriptsize}

The confidence interval of the calculated rates is now considered. The error on the model parameter E$_{bond}$  leads to an error factor of 2 (see Fig. \ref{fragCC_GCR} and Fig.~\ref{frag4CC_GCR}). The error on the (Q/Nc)$_{MF}$ parameter is difficult to estimate since only C$_{60}$ data are available to set it.  As it can be seen in Fig.~\ref{tauxQ_lifetime}, decreasing this parameter from the 4/60 value leads to a more drastic effect than increasing it. In other words, if PAHs are more robust that what we supposed ((Q/Nc)$_{MF}$ $>$ 4/60) it  affects slightly the results while if it is less robust it may  change it more. Accordingly, we estimate that the error from (Q/Nc)$_{MF}$ parameter may range in $\times$ 5 $\div$ 2. The rates are then  predicted with an estimated confidence interval of $\times$ 10 $\div$ 4.

\subsubsection{Impact on the gas phase}

CR destroy interstellar PAH and produce hydrocarbon fragments, so that any gas phase medium containing PAH will be seeded with long  hydrocarbon "chains". This is a continuous injection since the life times of PAH are very long. To address the gas phase abundance of these chains at steady state, the production rates given by the model have to be compared to the chain destruction rates.
\\
\\
\emph{- Diffuse medium }

 In the diffuse medium the destruction first occurs by photo-dissociation under UV field. Rates are  typically in the range of few 10$^{-9}$s$^{-1}$ under unshielded interstellar radiation field (ISRF)  \citep{2008CP....343..292V,1988ASSL..146...49V, Wakelam15}. If the chains contain hydrogen atoms, a few photons will be necessary to dehydrogenate before the photo-destruction of the skeleton occurs. Effective destruction rates by photon would then be lowered. Reactions of neutral species with ions (C$^+$) also cause carbon chains destruction \citep{2013ApJ...771...90C}. Here again, if the carbon chains contain hydrogen,  more than one reaction will be necessary to destroy the carbon chain skeleton. In a diffuse cloud at steady state,  most of the carbon is in the form of C$^+$.  With 500 H cm$^{-3}$, Av=0.2, T=110 K, $\zeta$ = 6.3 10$^{-17}$s$^{-1}$, the fractional abundance of C$^+$ calculated with the Nahoon code using the KIDA database and elementary composition of \cite{2011A&A...530A..61H} is  C$^+$/H = 1.68 10$^{-4}$ (for a total carbon to H ratio in gas phase of 1.70 10$^{-4}$). It corresponds to a density of 8.5 10$^{-2}$cm$^{-3}$. With an ion molecule reaction rates of a few 10$^{-9}$cm$^{3}$s$^{-1}$, the destruction rate by ion molecule reaction is then in the order of a few 10$^{-10}$s$^{-1}$. Other ions or atoms are in too low quantities or react too slowly to have a significant influence. The charged carbon chains in addition may be destroyed by dissociative recombination (DR). If hydrogen is present,  the carbon skeleton is preserved since H emission dominates and DR simply transforms the charged chains into neutrals.  In the diffuse medium, the electron fractional abundance  is approximately  equal to the C$^+$ abundance (8.5 10$^{-2}$cm$^{-3}$).   Adopting a DR rate of 10$^{-6}$cm$^{3}$s$^{-1}$ at 110 K \citep{Wakelam15}, the neutralization rates of the charged chains are then in the order of 10$^{-7}$s$^{-1}$.      
   
  Accordingly we will state that destruction rates in diffuse medium are 10$^{-9}$s$^{-1}$ for neutral chains and 10$^{-7}$s$^{-1}$ for charged ones (neutralization). Reaction of neutral chains with electrons leading to anion species are not considered because rates are still debated \citep{2013ApJ...774...97C, 2016PhRvL.117l3001K, A800490K}.

 The steady state abundances of chains for the different PAH  models and CR types, are shown in  Table  \ref{diffuse}. They were calculated using the destruction rates discussed above. For the production rates, we used the mean production values between a mass 5 and 10 with E$_{bond}$ = 6 eV (see for example  Fig.~ \ref{fragCC_GCR} to \ref{frag4CCx2_GCR}). We stated that 3 states of hydrogenation  of carbon chains are produced in equal quantities. The fractional abundance of C$_n$H$_m$ species (n=5-10; m=0-2)  reported in Table \ref{diffuse} have then been obtained by dividing the mean production rates by a factor three. 

The charged chains containing hydrogen are mostly transformed to neutral chains with conservation of the carbon skeleton due to hydrogen emission as we already discussed. The neutral fractional abundance reported in the Table~\ref{diffuse} have then be  increased by 50 $\%$.
 
\begin{table*}
\begin{center}
\small
\begin{tabular}{c c c c c c c }
\hline
 & & & & & & \\
 GM & [PAH]/[H] & CR & $\zeta$(s$^{-1}$) & Production rate (s$^{-1}$) & [C$_n$H$_m$]/[H] & [C$_n$H$_m^+$]/[H] \\
& & & & & & \\
\hline
& & & & & & \\
1 &  4.8(-7) & GCR S & 6.3(-17) & 5.9(-18) & 4.2(-15) & 2.8(-17) \\
  &                & GCR H & 1.6(-16)  & 2.8(-17) & 2.1(-14) & 1.4(-16) \\
  &                & E-1 S & 1.3(-15)  & 9.3(-17) & 6.6(-14)& 4.4(-16) \\
& & & & & & \\

2 &  1.2(-7) & GCR S & 6.3(-17)  & 8.4(-18) & 1.5(-15) & 1.0(-17)\\
  &                & GCR H & 1.6(-16)  & 4.2(-17) & 7.5(-15) & 5.0(-17)\\
  &                & E-1 S & 1.3(-15)  & 2.8(-16) & 5.0(-14) & 3.3(-16)\\
& & & & & & \\

3 &  1.2(-7) & GCR S & 6.3(-17)  & 2.2(-18) & 3.9(-16) & 2.6(-18) \\
  &                & GCR H & 1.6(-16)  & 1.1(-17) & 2.0(-15) & 1.3(-17) \\
  &                & E-1 S & 1.3(-15)  & 7.3(-17) & 1.3(-14) & 8.8(-17)\\
& & & & & & \\
\hline

\end{tabular}
\normalsize	    
\end{center}

\caption{\label{diffuse} Fractional abundances of carbon chains produced by CR in diffuse medium (500 H cm$^{-3}$, Av=0.2, T=110 K). A(B) = A$\times$10$^{B}$. First column: PAH  model type; second column: fractional  abundance of PAH,  third column:  CR type (S: with standard composition, H: with heavy enriched composition); fourth column: ionization rate of H$_2$,  fifth column:  production rates by CR in s$^{-1}$ of individual C$_n$H$_m$ (n = 5-10; m= 0-2) chains (see text); sixth column: fractional abundance of individual neutral chains;  last column:  fractional abundance of individual charged chains.}

\end{table*}

The fractional abundances of ionic species are two orders of magnitude lower than neutrals. This is due to the difference between  photo- dissociation  and DR rates.   Fractional abundances of the neutral chains  range from a few ~10$^{-16}$ to a few ~10$^{-14}$. The lowest abundance is obtained for the large fluffy PAH in a GCR spectrum with standard composition and the highest one for a small PAH in E-1 spectrum.

The destruction rates for a particular species may vary by factor of 5 with respect to the mean adopted values. Together with uncertainties on the productions rates, the abundances of Table \ref{diffuse} are predicted with an estimated confidence interval of $\times$ 50 $\div$ 20.
\\
\\
\emph{- Dense medium }

In quiescent dense cloud media (Av = 2 to 5)  for a particular combinations of age, temperature, and cloud density, PAH may be still free flying \citep{2008ApJ...680..371W}. If PAH are considered free,  the depletion on icy grains of molecular species (CO, H$_2$O, ...) is also assumed to be not achieved. The elementary abundances are the same that in the diffuse medium. ISRF photons cannot penetrate dense clouds. Neutral atomic species and the ions induced by CR are the main destructive species. Secondary UV photons induced by CR  are of less importance to destroy chains. The Table \ref{dense1} gives the list of neutral atomic species and ions destroying the chains. We do not consider, as in the case of the diffuse medium  the electron attachment. 
We displayed for each reaction a mean reaction rate over the size and hydrogenation of  neutral chains. Recent studies on C and O reactions with chains \citep{2014MNRAS.437..930L}, together with KIDA database 2014 \citep{Wakelam15}, were used. 

We performed steady state calculation with the Nahoon code \citep{Wakelam14} for various [O]/[C] ratio $>$ 1 to get the atomic and ion reactant concentrations. We did it for a medium of 2~10$^{4}$ H cm$^{-3}$, T= 10 K, Av=10, using the initial abundance of \cite{2011A&A...530A..61H}. Using these concentrations and the reaction rates, the destruction rates of neutral chains in dense medium are reported in the Table \ref{dense1}.

\begin{table*}
\begin{center}
\small
\begin{tabular}{ c c c c c }
\hline
 & & & & \\
Reactant & Reaction rate   & Destruction rate (s$^{-1}$)  & Destruction rate (s$^{-1}$)  & Destruction rate (s$^{-1}$) \\
         & (cm$^3$ s-1)  & \textsf{[O]/[C]=1.41}  & \textsf{[O]/[C]= 1.12} & \textsf{[O]/[C]= 1.06} \\
& & & & \\
\hline
& & & & \\
O    & 5.0(-11) & 1.8(-11) & 1.1(-11) & 7.0(-12)  \\
C    & 3.0(-10) & 1.3(-14) & 4.5(-12) & 7.2(-11)  \\
N    & 1.0(-13) & 2.9(-15) & 1.5(-12) & 1.2(-12)  \\
C+   & 1.0(-9)  & 1.1(-13) & 4.3(-13) & 4.2(-12)  \\
He+  & 2.5(-9)  & 2.1(-14) & 6.8(-15) & 7.0(-15)  \\
H$_3$+  & 2.0(-9)  & 4.2(-14) & 4.3(-14) & 4.6(-14)  \\
HCO+ & 1.5(-9)  & 1.6(-13) & 6.3(-14) & 7.2(-13)  \\
& & & & \\

\textbf{Total} &   -   & 1.8(-11) & 1.6(-11) & 8.3(-11) \\
\hline

\end{tabular}
\normalsize	    

\end{center}

\caption{\label{dense1} Destruction rates of neutral carbon chains in dense medium (2~10$^{4}$ H cm$^{-3}$, Av=10, T= 10 K).  First column: Reactant ; second column: reaction rate in cm$^{3}$ s$^{-1}$;  three last columns:  destruction  rates in s$^{-1}$ for [O]/[C] ratio respectively equal to 1.41, 1.12, 1.06. A(B) = A$\times$10$^{B}$.}
\end{table*}

For oxygen rich medium, the oxidation drives the destruction. For oxygen poor medium (unlikely to exist in ISM normal clouds), the reaction with carbon dominates. Keeping a ratio [O]/[C] well above 1, the total destruction rates is accordingly in the order of 2. 10$^{-11}$ s$^{-1}$ for neutral chains. For the charged species, the reactions with C and O are  of the ion molecule reaction type  with a typical reaction rate in the order of  10$^{-9}$ cm$^{-3}$ s$^{-1}$. It leads to a destruction rate of 7 10$^{-10}$ s$^{-1}$ for  [O]/[C]=1.41.  Destruction of charged chains are also resulting from DR. The fractional electron abundance, still for same [O]/[C] ratio, is calculated to be in the order of few 10$^{-8}$. It produces destruction rates in the order of  few 10$^{-10}$ s$^{-1}$.  Altogether the destruction rate of charged chains is in the order of 10$^{-9}$ s$^{-1}$. 
Like in the diffuse medium, charged chains are destroyed faster than neutral ones by around two orders of magnitude.   

  The fractional abundances of the chains in a dense medium are given in  Table \ref{dense}. The productions rates are the same that the ones used for the diffuse medium. We do not report on the E-1 spectrum  calculation because it fully transforms to GCR spectrum as soon as it is propagated within a medium with Av $>$ 1 \citep{Chabot16}. 

\begin{table*}
\begin{center}
\small
\begin{tabular}{c c c c c c c}
\hline
 & & & & & &  \\
 GM & [PAH]/[H] & CR & $\zeta$ & Production rate & [CnHm]/[H] & [CnHm$^+$]/[H] \\
 & & & (s$^{-1}$) & (s$^{-1}$) & &  \\
& & & & & &  \\
\hline
& & & & & &  \\
1 &  4.8(-7) & GCR S & 6.3(-17) & 5.9(-18) & 2.0(-13) & 2.8(-15) \\

  &                & GCR H & 1.6(-16)  & 2.8(-17) & 1.0(-12) & 1.4(-14)\\
& & & & & &  \\
2 &  1.2(-7) & GCR S & 6.3(-17) & 8.4(-18) & 7.6(-14) & 1.0(-15)\\
  &                & GCR H & 1.6(-16)  & 4.2(-17) & 3.8(-13) & 5.0(-15)\\
& & & & & &  \\
3 &  1.2(-7) & GCR S & 6.3(-17)  & 2.2(-18) & 2.0(-14) & 2.6(-16) \\
  &                & GCR H & 1.6(-16)  & 1.1(-17) & 1.0(-13) & 1.3(-15) \\
& & & & & &  \\
\hline

\end{tabular}
\normalsize	    
\end{center}

\caption{\label{dense} Same legend as the table \ref{diffuse} for a dense medium (2 ~10$^{4}$ H cm$^{-3}$, Av=10, T=10~K).}
\end{table*}

The fractional abundance of chains in dense medium is ranging between 1.~10$^{-12}$ and 2.~10$^{-14}$. With the adopted fractional abundance of PAH (PAH/H $ \approx $ 1.~10$^{-7}$), the carbon chains abundance per PAH is found to range between 1.~10$^{-5}$  and 2.~10$^{-7}$. 
       
Uncertainties on these calculated abundances are similar to the ones for the diffuse medium. i.e $\times$ 50 $\div$ 20. They are coming from model errors and from errors on the adopted destruction rates.
\\
\\ 
\emph{- Confrontation to observations and other chain production mechanisms }

The predicted carbon chains abundances are below the present observation capabilities. By consequence, the coulomb explosion of PAH it not involved in any of the many reported detections of carbon chains, namely in circumstellar envelopes of carbon rich asymptotic giant branch stars  \citep{1996A&A...309L..27C}, in cold dense (quiescent) clouds \citep{1983ApJ...270..589B, 1997ApJ...483L..61B, 2010ApJ...718L..49S, 2016ApJ...824..136L, 2009ApJ...699..585H}, and in protostar cores \citep{Sakai2008}.           
In the diffuse medium, upper limits  for some long chains have been established \citep{2008A&A...486..493L, Liszt2012}. These limits are higher (compatible) with the present results.

The steady state abundances of carbon chains from a pure gas phase chemistry are very weak. For example, in dense cloud, using  Nahoon  and the KIDA database, $\approx$10$^{-18}$ for C$_8$H and in diffuse medium $\approx$5.10$^{-16}$ for C$_5$H. It is many order of magnitude lower than the carbon chains abundances produced by the coulomb explosion of PAH.  

\section{Conclusions}

In this work, we modeled the CR induced  multi-ionization cross sections of interstellar PAHs.
We calculated the carbon chains production resulting from the subsequent coulomb explosion taking place for high Q fold ionization of the PAH. The calculations were found in  reasonable  agreement with existing experiments performed on the C$_{60}$ molecule,  small hydrocarbons, and PAH. Additional experiments would help in further constraining the details of the model.      

With double differential cross sections in energy and projectile atomic number,  the carbon chains production rates using  different CR types and for different size and compactness of PAH models  were computed. The size of the carbon chains resulting from the coulomb explosion  extends up to 10 to 15($\pm$3) carbon atoms, depending on the size, compactness and hydrogenation degree of the PAHs. Their production rates are weakly depending on their sizes. The carbon chains production rates range from a few to many tens of percents of the H$_2$ ionization rate, depending on the PAH structure and the adopted CR type (i.e. Galactic standard spectrum or closer to a local CR source with a steeper low energy flux distribution). 

In the diffuse medium under GCR spectrum, for a depletion of carbon in PAH of $\approx$ 10$\%$ of the available cosmic carbon, the  estimated chain steady state fractional abundances range from  ~10$^{-15}$ to ~10$^{-14}$. In dense clouds, shielded from external UV at moderate Av, they range around ~10$^{-13}$. The uncertainties are in the order of $\times$50 $\div$20. They are  due to the approximations on the model, the poor knowledge of the exact structure of the PAH, and the  determination of destruction rates. For the long carbon chains, with a number of carbon greater than 6, these abundances are 10$^{4-6}$ larger than pure gas phase steady states abundance in both diffuse and dense medium.   

Observationally, the column density sensitivity capabilities with present astrophysical observatories  make the detection of this process certainly difficult. Nevertheless in the surrounding environment of PAH, the calculations show that there is around 10$^{-6}$ - 10$^{-7}$ of carbon chains per PAH  due to the CR coulomb explosion.

\begin{acknowledgements}
This work is performed within the framework of the P2IO LabEx program « Evolution de la mati\`ere du milieu interstellaire aux exoplan\'etes avec le JWST ». The authors acknowledge the CNRS programs  « Physique et Chimie du Milieu Interstellaire » (PCMI) and « Edifices Mol\'eculaires Isol\'es et Environn\'es » (GDR-EMIE). We thank Cyril Falvo, Vincent Tatischeff and Serge Martin for help and discussions. 
\end{acknowledgements}

\bibliographystyle{aasjournal}
\bibliography{CE}


\end{document}